\documentclass[preprint,showpacs,preprintnumbers,amsmath,amssymb]{revtex4}
\newcommand{\newwidth}{0.675\textwidth}
\newcommand{\newheight}{0.45\textwidth}
\newcommand{\newwidthprime}{0.225\textwidth}
\newcommand{\newheightprime}{0.30\textwidth}

\usepackage{graphicx,verbatim}
\usepackage{dcolumn}
\usepackage{bm}
\usepackage{sidecap}
\graphicspath{{/user/rhaertle/epsfiles/Propplots/},{/user/rhaertle/Berlin08/},
{/user/rhaertle/epsfiles/Rate-Report/},{/user/rhaertle/epsfiles/RateCoul/},
{/user/rhaertle/epsfiles/Benesch/},{/user/rhaertle/PaperI/},{/user/rhaertle/PaperII/},
{/user/rhaertle/epsfiles/Cizek/}}

\newcommand{\tr}[1]{\text{Tr}\lbrace #1 \rbrace}
\newcommand{\trR}[1]{\text{Tr}_{\text{B}}\lbrace #1 \rbrace}
\newcommand{\trRprime}[1]{\text{Tr}_{\text{B}'}\lbrace #1 \rbrace}

\begin{document}

\title{
Resonant Electron Transport in Single-Molecule Junctions:
Vibrational Excitation, Rectification, Negative Differential Resistance and Local Cooling
}

\author{R.\ H\"artle}
\author{M.\ Thoss}
\affiliation{
Institut f\"ur Theoretische Physik and Interdisziplin\"ares Zentrum f\"ur Molekulare 
Materialien, \\
Friedrich-Alexander-Universit\"at Erlangen-N\"urnberg,\\ 
Staudtstr.\,7/B2, D-91058 Erlangen, Germany
}

\date{\today}

\begin{abstract}

Vibronic effects in resonant electron transport through single-molecule
junctions are analyzed. The study is based on generic models for molecular
junctions, which include electronic states on the molecular
bridge that are vibrationally coupled and exhibit Coulomb
interaction. The transport calculations employ a master equation approach. 
The results, obtained for a series of
models with increasing complexity, show a
multitude of interesting transport phenomena, including vibrational
excitation, rectification, negative differential resistance (NDR) as well as
local cooling.  While some of these phenomena have been observed or
proposed before, the present analysis extends previous studies and allows a
more detailed understanding of the underlying transport mechanisms.
In particular, it is shown that many of the observed phenomena can only be
explained if  electron-hole pair  creation processes at the molecule-lead 
interface are taken into account.
Furthermore, vibronic effects in sytems with multiple electronic states and
their role for the stability of molecular junctions are
analyzed.
\end{abstract}


\maketitle

\section{Introduction}

For more than a decade, molecular electronics 
\cite{Nitzan01,Cuniberti05,Selzer06,Tao2006,Venkataraman06,Chen07,GalpScience08,Molen2010,cuevasscheer2010} has 
been a very active and challenging field of research. One of the basic ideas is to 
exploit the diversity of molecules and the possibilities of modern synthesis
to design molecular systems with specific functions for nanoscale electronic devices.
Another motivation is the possibility to investigate single molecules under 
controllable nonequilibrium conditions. 
Various techniques, including mechanically 
controlled break junctions \cite{Reed97,Reichert02,Smit02,Boehler04,Elbing05,Martin2010}, 
electro-migrated molecular junctions \cite{Natelson04,Sapmaz05,Sapmaz06,Leon2008,Huettel2009,Osorio2010}, 
scanning tunneling microscopy \cite{Ho98,WuNazinHo04,Ogawa07,Schulze08,Pump08,Tao2010}, 
and very recently, on-wire lithography in combination with \emph{in-situ} 'click chemistry' 
\cite{Mayor2009,Braunschweig2009}, have been employed to contact a single molecule with two 
macroscopic electrodes. Once such a molecular junction is established, external electric fields,
 either a bias or a gate voltage  \cite{Sapmaz05,Song2009,Osorio2010,Martin2010}, can be used 
to investigate the conductance of a single molecule. A variety of interesting transport phenomena
 have been proposed and experimentally observed \cite{Selzer06,Tao2006,Yeganeh2007,Chen07,Molen2010,cuevasscheer2010}, including, 
\emph{e.g.}, switching behavior \cite{Riel2006,Choi,Evers2009,Benesch2009}, rectification 
\cite{WuNazinHo04,Elbing05,DiezPerez09} and negative differential resistance 
\cite{Gaudioso00,LeRoy,Pop2005,Sapmaz05,Sapmaz06,Osorio2010}.

However, a detailed understanding of the experimental results, especially in the resonant 
transport regime, where electrons may populate an intermediate state on the bridging molecule, 
has not been achieved yet. One of the interesting and challenging aspects of electron transport in
molecular junctions is the intricate interplay between 
the electronic and vibrational degrees of freedom. 
Because of the small size of molecules, the charging of the molecular bridge 
is often accompanied
by significant changes of the nuclear geometry that indicate
strong coupling between electronic and vibrational degrees of freedom.
As a consequence,  the vibrational 
modes of a molecular junction can be highly excited resulting in significant
nonequilibrium effects \cite{Hartle,Hartle09,Romano10,Hartle2010}. 
This aspect distinguishes single-molecule junctions from traditional  quantum dot systems
\cite{Reed1988,Holleitner2001,Kubala2002,Datta2007}. 
Vibrational signatures indicating strong vibronic coupling as well as strong excitation 
of vibrational modes were identified for a number of molecular junctions
\cite{WuNazinHo04,LeRoy,Natelson04,Pasupathy05,Sapmaz06,Thijssen06,Parks07,Boehler07,Leon2008,Huettel2009,Tao2010,Ballmann2010,Jewell2010,Osorio2010}. 
Novel experimental techniques based on measuring the force 
needed to break  a junction \cite{Huang2006} or employing  Raman spectroscopy \cite{Natelson2008,Ioffe08}
allow a characterization of the current-induced
vibrational nonequilibrium state of a single-molecule junction. 
These data complement the information carried by the 
respective current-voltage characteristics.

Various theoretical approaches have been employed to describe vibrationally coupled
 electron transport through single molecules. While scattering theory approaches 
\cite{Cizek04,Toroker07,Zimbovskaya09,Seidemann10} can be used to address the 
regime of strong molecule-lead coupling, nonequilibrium Green's function approaches 
\cite{Flensberg03,Mitra04,Galperin06,Ryndyk06,Frederikesen07b,Tahir2008,Hartle,Stafford09,Hartle09} 
additionally allow a non-perturbative description of the associated nonequilibrium 
state of such a junction, especially with respect to the vibrational degrees of freedom.
 Numerically exact methods, based on path integrals \cite{Muehlbacher08,Thorwart2008} 
or multiconfigurational wave-function methods \cite{Wang09}, provide valuable 
insights and benchmarks for specific model systems and problems that may not be addressed
 by perturbation theory or other approximative schemes. Master equation approaches 
\cite{May02,Mitra04,Lehmann04,Pedersen05,Harbola2006,Zazunov06,Siddiqui,Timm08,May08,May08b,Leijnse09,Esposito09,Esposito2010},
 although perturbative with respect to the coupling between the molecule and the leads,
 have been proven to be very powerful, as they are capable of describing all interactions
 on the molecular bridge accurately and,  for simple model systems, very efficiently.
In this work, we employ a master equation approach that is based on a second order expansion
 in the molecule-lead coupling \cite{Mitra04,Lehmann04,Harbola2006}. Master equation approaches 
that take into account higher-order effects with respect to this coupling have already 
been employed \cite{May02,Pedersen05,Leijnse09,Esposito09,Esposito2010}. The corresponding 
higher-order transport processes 
\cite{Koenig96,Paaske2005,Galperin2007,Lueffe,Hartle,Leijnse2009,Han2010}, however, 
are beyond the scope of this work, where we focus on the resonant transport regime.

In this article, we analyze nonequilibrium transport phenomena induced by
electronic-vibrational coupling in molecular junctions. To this end, we consider generic
models for vibrationally coupled electron transport including electronic states 
on the molecular bridge that are vibrationally coupled and exhibit Coulomb interaction.
The results show a multitude of interesting phenomena that extend previous
studies and may facilitate the interpretation of experimental results. 
In particular, it is found that  the current-induced vibrational excitation
in transport through a molecular bridge with a single electronic state
increases significantly with increasing bias voltage and/or decreasing electronic-vibrational coupling.
We show that this phenomenon is caused by electron-hole pair creation processes 
\cite{Ueba2002,Ueba2002b,noteonelholepaircreation}, which to the best of our knowledge have
not been considered in detail in this context before. 
Further analysis shows that electron-hole pair creation processes can also
explain the strong enhancement of  vibrational rectification effects in
situations where the vibrational degree of freedom acquires a highly excited nonequilibrium state.
This complements the analysis of the phenomenon of vibrational rectification
given in Ref.\ \onlinecite{Flensberg03}. 
Electronic-vibrational coupling may also cause vibrationally induced NDR
effects \cite{Schoeller01,Koch05b,Zazunov06,Leijnse09}. In this work, we reinterpret the NDR-mechanism outlined in Refs.\ \onlinecite{Schoeller01,Zazunov06} in terms 
of pair creation processes. Furthermore, a novel mechanism
 for vibrationally induced NDR is discussed, which, in contrast to earlier studies 
\cite{Hettler2002,Hettler2003,Datta2007}, extends over a broad range of bias voltages.

In molecular junctions, where multiple electronic states participate in
the charge transport, a number of additional vibronic
processes have to be considered. In particular, as we have shown recently \cite{Hartle09}, 
higher-lying electronic states 
facilitate resonant absorption processes that may deexcite the vibrational degrees of
 freedom. This mechanism reduces the current-induced vibrational excitation of a
 molecular junction and results in local cooling \cite{Hartle09,Romano10}.
In the present paper, we give a detailed analysis of  effects due to multiple electronic states. 
In this context, we also show that repulsive Coulomb interactions
may further enhance the stability of a molecular junction
(cf.\ \ref{SecCoulombCooling}). We refer to this phenomenon as 'Coulomb
Cooling'.   Since polyatomic molecules typically include numerous active
vibrational modes and often exhibit multiple closely lying
electronic states, these phenomena are expected to be of
relevance for most molecular junctions.

The article is organized as follows. In Section \ref{Hamsec} we introduce the model
 Hamiltonian used to describe electron transport through a single-molecule junction.
 The derivation of the master equation approach and expressions for the observables of
 interest, in particular, current-voltage characteristics and the average vibrational excitation of a molecular
 junction, are outlined in Secs.\ \ref{gentheo} and \ref{currentandvibex}. Explicit 
formulas adapted to the specific model systems are detailed in 
appendices \ref{specMEonestate} and \ref{specMEtwostates}.
The role of coherences is elucidated in appendix \ref{effectofvibrationalcoherences}.
Sec.\ \ref{secResults} comprises  numerical results and a discussion of the 
different transport phenomena.
Thereby, we consider transport phenomena that involve a single electronic state 
(Sec.\ \ref{secResonestate}) and two electronic states (Sec.\ \ref{restwostates}).
Besides the basic transport mechanisms that are discussed in Sec.\ \ref{basmech} and 
\ref{ResonantAbsorption}, we study vibrationally induced rectification in Secs. 
\ref{secvibrect} and \ref{ResonantAbsorptionAsym} as well as vibrationally induced negative 
differential resistance in Secs.\ \ref{secvibNDR} and
\ref{twostateNDR}. Vibrational excitation processes in junctions with two
electronic states and their influence on the stability of a molecular junction
are analyzed  in Secs.\ \ref{ResonantAbsorption} to \ref{SecCoulombCooling}.
Throughout the article we use units where $\hbar=1$.

\section{Theory}

\subsection{Model Hamiltonian}
\label{Hamsec}

We consider electron transport through a single molecule that is
bound to two metal leads. Such a molecular junction is described by a set of discrete 
electronic states, which are localized on the molecular bridge (M) and interact
with a continuum of electronic states in the left (L) and the right (R)
lead, respectively. The corresponding model Hamiltonian is given by 
\begin{eqnarray}
\label{hel}
H_{\text{el}} &=& \sum_{i\in\text{M}} \epsilon_{i} c_{i}^{\dagger}c_{i} 
+ \sum_{k\in\text{L,R}} \epsilon_{k} c_{k}^{\dagger}c_{k} \\
&& + \sum_{i<j\in\text{M}} U_{ij} (c_{i}^{\dagger}c_{i}-\delta_{i})
 (c_{j}^{\dagger}c_{j}-\delta_{j}) \nonumber \\
&& + \sum_{k\in\text{L,R};i\in\text{M}} ( V_{ki} c_{k}^{\dagger}c_{i} + \text{h.c.} ). \nonumber
\end{eqnarray}
Thereby, $\epsilon_{k}$ denote the energies of the
lead states with corresponding creation and annihilation operators
$c_{k}^{\dagger}$ and $c_{k}$. Likewise, $\epsilon_{i}$ is 
the energy of the $i$th electronic state on the molecular 
bridge, which is addressed by creation and annihilation 
operators $c_{i}^{\dagger}$ and $c_{i}$.
The coupling matrix elements $V_{ki}$ characterize the 
strength of the interaction  between the electronic states of 
the molecular bridge and the leads and determine the so-called level-width 
functions 
$\Gamma_{K,ij}(\epsilon)=2\pi\sum_{k\in K} V_{ki}^{*} V_{kj}
\delta(\epsilon-\epsilon_{k})$ ($K$=L,R). 

Additional charging energies, due to Coulomb interactions, are 
accounted for by Hubbard-like electron-electron interaction terms, 
$U_{ij}(c_{i}^{\dagger}c_{i}-\delta_{i}) (c_{j}^{\dagger}c_{j}-\delta_{j})$.
Thereby, the parameters $\delta_i$ distinguish  states that are occupied ($\delta_{i}=1$)
or unoccupied ($\delta_{i}=0$) in the molecular system at equilibrium. 
While in the present study we use a generic model, the first-principles determination
of the parameters requires the introduction of a reference system
\cite{Benesch08}. In the present paper, we consider the neutral molecule in
its electronic ground state as a reference system.
As a result, for an electronic state 
above the Fermi-level ($\delta_{i}=0$), 
the associated energy $\epsilon_{i}$ denotes 
the energy required to add 
an electron to the $i$th electronic state of the reference system. 
For an electronic state
below the Fermi-level ($\delta_{i}=1$), $\epsilon_{i}$ denotes 
the energy required to remove 
an electron from state $i$, 
in accordance with Koopmans' theorem.
The Fermi energy of the leads is set to $\epsilon_{\text{F}}=0$\,eV.

Upon transmission of electrons, the molecular bridge may be vibrationally excited. 
We describe the vibrational degrees of freedom of a molecular junction within
the harmonic approximation,
\begin{eqnarray}
\label{Hvib}
H_{\text{vib}} &=& \sum_{\alpha} \Omega_{\alpha} a_{\alpha}^{\dagger}a_{\alpha} 
+ \sum_{i\in\text{M};\alpha} \lambda_{i\alpha} Q_{\alpha} (c_{i}^{\dagger}c_{i}-\delta_{i}),
\end{eqnarray}
where the operator $a^{\dagger}_{\alpha}$ denotes the creation operator of the 
$\alpha$th oscillator with frequency $\Omega_{\alpha}$.
The coupling between the electronic and the vibrational degrees of 
freedom is assumed to be linear in both the vibrational displacements 
$Q_{\alpha}=a_{\alpha}+a_{\alpha}^{\dagger}$ and the electron (or hole) 
densities $(c_{i}^{\dagger}c_{i}-\delta_{i})$ \cite{Cederbaum74,Benesch08,Han2010}. The respective 
coupling strengths are denoted by $\lambda_{i\alpha}$. 
Because we employ the normal modes of the ground-state of the neutral molecule, there is no coupling between the 
electronic states and the normal modes of the molecular junction in this state. This is imposed 
in the electronic-vibrational coupling term by the parameters $\delta_{i}$.
The Hamilton operator of the overall system is given by the sum
\begin{eqnarray}
H &=& H_{\text{el}} + H_{\text{vib}}. 
\end{eqnarray}

In the limit of vanishing molecule-lead coupling, $V_{ki}\rightarrow0$, 
the Hamiltonian $H$ can be diagonalized by the
 small polaron transformation \cite{Mahan81,Koenig96,Mitra04,Hartle}
\begin{eqnarray}
\label{transformedHamiltonian}
\overline{H} &=& \text{e}^{S} H \text{e}^{-S} = \overline{H}_{\text{S}} 
+ \overline{H}_{\text{B}} + \overline{H}_{\text{SB}}, \\
\overline{H}_{\text{S}}&=& \sum_{i} \overline{\epsilon}_{i} c_{i}^{\dagger}c_{i} 
+ \sum_{\alpha} \Omega_{\alpha} a^{\dagger}_{\alpha}a_{\alpha} \nonumber\\
&& + \sum_{i<j} \overline{U}_{ij} (c_{i}^{\dagger}c_{i}-\delta_{i})
(c^{\dagger}_{j}c_{j}-\delta_{j}), \nonumber\\
\overline{H}_{\text{B}}&=& \sum_{k} \epsilon_{k} c_{k}^{\dagger}c_{k},  \nonumber\\
\overline{H}_{\text{SB}} &=& \sum_{ki} ( V_{ki} X_{i} 
c_{k}^{\dagger}c_{i} + \text{h.c.} ),  \nonumber
\end{eqnarray}
with 
\begin{eqnarray}
S &=& \sum_{i\alpha} \frac{\lambda_{i\alpha}}{\Omega_{\alpha}}  ( c^{\dagger}_{i}c_{i} 
- \delta_{i}  ) 
 (a^{\dagger}_{\alpha}-a_{\alpha}), \\
X_{i} &=& \text{exp}[\sum_{\alpha}\frac{\lambda_{i\alpha}}{\Omega_{\alpha}}
(a_{\alpha}-a_{\alpha}^{\dagger})].
\end{eqnarray}
Thereby, we have partitioned the transformed Hamiltonian $\overline{H}$ in three parts, 
$\overline{H}=\overline{H}_{\text{S}}+\overline{H}_{\text{B}}+\overline{H}_{\text{SB}}$,
with $\overline{H}_{\text{S}}$ representing the molecular bridge, 
$\overline{H}_{\text{B}}$  the leads and $\overline{H}_{\text{SB}}$ 
describing the interactions between the molecular bridge and the leads. 
Due to the small-polaron transformation, there is no explicit electronic-vibrational 
coupling in $\overline{H}_{\text{S}}$. However, electronic-vibrational 
coupling appears in the transformed Hamiltonian $\overline{H}$ at three different 
places:
\begin{itemize}
\item{in the polaron-shifted energies: 
${\overline{\epsilon}_{i}=\epsilon_{i}+(2\delta_{i}-1)\sum_{\alpha}
(\lambda_{i\alpha}^{2}/\Omega_{\alpha})}$,}
\item{in additional electron-electron interactions, 
which shift the original electron-electron interaction terms:
 ${\overline{U}_{ij}=U_{ij}-2\sum_{\alpha}(\lambda_{i\alpha}
\lambda_{j\alpha}/\Omega_{\alpha})}$,}
\item{and in the molecule-lead coupling term $\overline{H}_{\text{SB}}$ 
that is renormalized by the shift operators $X_{i}$.}
\end{itemize}

\subsection{Master equation approach}
\label{gentheo}

Density matrices have been proven to be a powerful tool in describing 
quantum-mechanical systems \cite{Blum,Louisell,Davidson,Weiss93,May04}.
Once the density matrix $\varrho$ of a given system is known, 
all observables $O$ of that system can be obtained from the trace
\begin{eqnarray}
\label{basicexpvalue}
\langle O \rangle = \tr{ \varrho O} = \sum_{a} \left\langle a \right\vert 
\varrho O \left\vert a \right\rangle = \sum_{ab} \varrho_{ab} O_{ba}.
\end{eqnarray}
Thereby, the elements of the density matrix are given as 
$\varrho_{ab} = \left\langle a \vert \varrho  \vert b \right\rangle$,
where $\vert a\rangle$ and $\vert b\rangle$ are elements of a complete 
set of orthonormal basis functions that span the Hilbert space of the overall system.
Since Eq.\ (\ref{basicexpvalue}) is invariant 
under the small polaron transformation, we consider in the following $\overline{H}$ as 
the Hamiltonian of the system.

The time-evolution of a density matrix $\varrho(t)$
 is determined by the Liouville - von Neumann equation:
\begin{eqnarray}
\label{vonNeumann}
\frac{\partial\varrho(t)}{\partial t} &=& -i \left[\overline{H},\varrho(t) \right]
 \equiv -i \mathcal{L}\varrho(t),\\
\varrho(t)&=&\text{e}^{-i\mathcal{L}t}\varrho(0),
\end{eqnarray}
where the initial state at time $t=0$ is encoded in the respective density matrix $\varrho(0)$. 
Here, $\mathcal{L}$ denotes the Liouville operator 
$\mathcal{L}\ \varrho\equiv\left[\overline{H},\ \varrho \right]$.

To describe an open quantum  system such as a single molecule coupled
to a reservoir of electrons (in the left 
and the right electrode), it is expedient to employ the reduced density matrix of
this system $\rho$, which is obtained by taking the trace over the degrees of freedom
of the reservoirs (or baths (B)),
\begin{eqnarray}
\rho(t) &=&\trR{\varrho(t)}.
\end{eqnarray}
Formally, this can be achieved by the projection operator
\begin{eqnarray}
P\varrho(t)&=&\rho_{\text{B}}\trR{\varrho(t)}\equiv\rho_{\text{B}}\rho(t),
\end{eqnarray}
and its orthogonal complement $Q=1-P$. 
Here, $\rho_{\text{B}}$
denotes the thermal equilibrium density matrix of the reservoir
\begin{eqnarray}
\rho_{\text{B}} &=& \mathcal{Z}^{-1} \text{e}^{-\beta \overline{H}_{\text{B}}},\quad
 \mathcal{Z}=\trR{\text{e}^{-\beta \overline{H}_{\text{B}}}}.
\end{eqnarray}

Assuming a factorized initial condition
$\varrho(0)=\rho(0)\rho_{\text{B}}$, 
the equation of motion for the
reduced density matrix is given by the Nakajima-Zwanzig equation \cite{Nakajima,Zwanzig}
\begin{eqnarray}
\label{NakajimaZwanzigEqu}
\frac{\partial}{\partial t}P\varrho(t) &=& -iP\mathcal{L}P \varrho(t)  -
 \int_{0}^{t} \text{d}\tau\, P\mathcal{L} \text{e}^{-iQ\mathcal{L}\tau} 
Q\mathcal{L} P\varrho(t-\tau), \nonumber\\
\end{eqnarray}
which represents a formally exact equation.
For practical applications, approximations are required to solve this equation
of motion.
In the present context, we assume that the molecule-lead coupling,
$\overline{H}_{\text{SB}}$,  is weak. Employing a second order expansion in
$\overline{H}_{\text{SB}}$ and
the condition $\text{tr}_{\text{B}}\lbrace \overline{H}_{\text{SB}} \rho_{\text{B}}
\rbrace=0$, 
the Nakajima-Zwanzig equation can be simplified to the master equation  \cite{May02,Egorova03,Mitra04,Lehmann04}
\begin{eqnarray}
\label{secondorderME}
\frac{\partial}{\partial t}\rho(t) &=& -i\left[ \overline{H}_{\text{S}} , \rho(t) \right] \\
&&  - \int_{0}^{\infty} \text{d}\tau\, \text{tr}_{\text{B}}\lbrace \left[ \overline{H}_{\text{SB}}, 
\left[ \overline{H}_{\text{SB}}(\tau), \rho(t) \rho_{\text{B}} \right] \right]
\rbrace, \nonumber
\end{eqnarray}
with
\begin{eqnarray}
\overline{H}_{\text{SB}}(\tau) &=& \text{e}^{-i(\overline{H}_{\text{S}}+\overline{H}_{\text{B}})\tau}
 \overline{H}_{\text{SB}} \text{e}^{i(\overline{H}_{\text{S}}+\overline{H}_{\text{B}})\tau}.
\end{eqnarray}
To obtain the time-local master equation described
by Eq.\ (\ref{secondorderME}),
we have, furthermore, employed the Markov approximation, which
involves the shift of the integration limit $\int_{0}^{t}\rightarrow\int_{0}^{\infty}$
and the replacement
\begin{eqnarray}
\rho(t-\tau) \approx \text{e}^{i\mathcal{L}_{\text{S}}\tau} \rho(t). 
\end{eqnarray}
The latter approximation is in line with the second order expansion in
$\overline{H}_{\text{SB}}$.
Due to the shift in the integration limit, the master equation  (\ref{secondorderME}) is only valid for
times longer than the correlation time of the bath 
\cite{Egorova03,Volkovich2008,noteonNonMarkovianEffects}. This is the case for
the applications to steady-state transport considered in this paper, where
only the long-time limit $\rho(t\rightarrow\infty)\equiv\rho$ is required. 
Taken in the basis of eigenstates of the system Hamiltonian $\overline{H}_\text{S}$, the master
equation (\ref{secondorderME}) corresponds to the Redfield equation \cite{Redfield1965,Blum,May04}.

In the steady state transport regime, 
the above equation of motion becomes an algebraic set of
equations
\begin{eqnarray}
\label{genfinalME}
0 &=& -i \left[ \overline{H}_{\text{S}} , \rho \right] \\
&& - \int_{0}^{\infty} 
\text{d}\tau\, \text{tr}_{\text{B}}\lbrace \left[ \overline{H}_{\text{SB}} ,
 \left[ \overline{H}_{\text{SB}}(\tau), \rho \rho_{\text{B}} \right] \right] 
\rbrace , \nonumber
\end{eqnarray}
which can be solved 
by standard linear algebra techniques.
Thereby, the normalization constraint $\text{tr}_{\text{S}}\lbrace \rho \rbrace=1$ ensures
a unique solution. 

As discussed in the Introduction, the master equation (\ref{secondorderME})
  as well as its steady-state form, Eq.\  (\ref{genfinalME}), is valid for small
molecule-lead coupling. Due to the neglect of terms of higher order in the system-reservoir coupling, it cannot
describe tunneling in the non-resonant transport regime and misses the broadening of
resonances due to this coupling. Except for these deficiencies,
however, it provides a rather
accurate description of vibrationally coupled electron transport in the
resonant transport regime considered in this work. This has been demonstrated
recently by comparison with results of
nonequilibrium Green's function (NEGF) methods  \cite{Hartle09}.
Test calculations show that all effects discussed in this work
are also obtained with  NEGF methods for vibrationally coupled resonant electron transport
\cite{Galperin06,Hartle,Hartle09}.

The basis functions, which we use to evaluate the reduced density matrix $\rho$ 
and the master equation Eq.\ (\ref{genfinalME}), are products of basis functions 
$\vert a \rangle\vert \nu \rangle$ that span the subspace of the electronic 
$\vert a \rangle$ and the vibrational  degrees of freedom $\vert \nu \rangle$, respectively. 
Thereby, the electronic basis functions are given in the occupation number 
representation \emph{i.e.}\ $\vert a \rangle=\vert n_{1}n_{2}.. \rangle$, where 
$n_{i}\in\lbrace0,1\rbrace$ denotes the population of the $i$th electronic state. 
Throughout this article we consider a single vibrational mode with frequency $\Omega$. 
Hence, we represent the vibrational basis functions by harmonic oscillator
basis functions $\vert\nu\rangle$, where 
$\nu\in\mathbb{N}_{0}$ stands for the excitation number of the vibrational mode. 
Thus, the coefficients of the reduced density matrix can be written as
\begin{eqnarray}
 \rho_{a,a'}^{\nu_{1}\nu_{2}} \equiv \langle a \vert \rho^{\nu_{1}\nu_{2}} 
\vert a' \rangle \equiv \langle a \vert \langle \nu_{1} \vert \rho \vert \nu_{2} 
\rangle \vert a' \rangle,
\end{eqnarray}
where upper case indices refer to states of the vibrational mode and lower case 
indices represent the electronic part of the respective Hilbert space.

Evaluating Eq.\ (\ref{genfinalME}) first between  vibrational states 
$\langle\nu_{1}\vert$ and $\vert\nu_{2}\rangle$, we obtain
the equation
\begin{eqnarray}
\label{specME}
&& - i \langle\nu_{1}\vert \left[\overline{H}_{\text{S}},\rho\right] \vert\nu_{2}\rangle = \\
&&\hspace{-0.3cm}\phantom{+} \pi \sum_{kij\nu_{3}\nu_{4}} V_{ki} V_{kj}^{*}  
 f_{k} X_{i,\nu_{1}\nu_{3}} X_{j,\nu_{3}\nu_{4}}^{\dagger} c_{i}  c_{j}^{\dagger} 
 \delta(E_{j,\nu_{3}\nu_{4}}) \rho^{\nu_{4}\nu_{2}} \nonumber\\
&&\hspace{-0.3cm}- \pi \sum_{kij\nu_{3}\nu_{4}} V_{ki} V_{kj}^{*}
(1-f_{k}) X_{i,\nu_{1}\nu_{3}} X_{j,\nu_{4}\nu_{2}}^{\dagger} 
 c_{i} \rho^{\nu_{3}\nu_{4}} c_{j}^{\dagger}  \delta(E_{j,\nu_{4}\nu_{2}})
    \nonumber\\
&&\hspace{-0.3cm}+ \pi \sum_{kij\nu_{3}\nu_{4}} V_{ki} V_{kj}^{*}   f_{k} 
X_{i,\nu_{3}\nu_{4}} X_{j,\nu_{4}\nu_{2}}^{\dagger}  \rho^{\nu_{1}\nu_{3}} c_{i} 
 \delta(E_{i,\nu_{4}\nu_{3}})  c_{j}^{\dagger} \nonumber\\
&&\hspace{-0.3cm}- \pi \sum_{kij\nu_{3}\nu_{4}} V_{ki} V_{kj}^{*} (1-f_{k}) 
X_{i,\nu_{1}\nu_{3}} X_{j,\nu_{4}\nu_{2}}^{\dagger}   c_{i}  
 \delta(E_{i,\nu_{3}\nu_{1}}) \rho^{\nu_{3}\nu_{4}} c_{j}^{\dagger}
\nonumber\\
&&\hspace{-0.3cm} + \pi \sum_{kij\nu_{3}\nu_{4}} V_{ki}^{*} V_{kj}    (1-f_{k}) 
X^{\dagger}_{i,\nu_{1}\nu_{3}} X_{j,\nu_{3}\nu_{4}}  c_{i}^{\dagger}  c_{j} 
 \delta(E_{j,\nu_{4}\nu_{3}}) \rho^{\nu_{4}\nu_{2}} \nonumber\\
&&\hspace{-0.3cm} - \pi \sum_{kij\nu_{3}\nu_{4}} V_{ki}^{*} V_{kj} f_{k} 
X^{\dagger}_{i,\nu_{1}\nu_{3}} X_{j,\nu_{4}\nu_{2}}  c_{i}^{\dagger}  \rho^{\nu_{3}\nu_{4}} c_{j} 
 \delta(E_{j,\nu_{2}\nu_{4}})  \nonumber\\
&&\hspace{-0.3cm} + \pi \sum_{kij\nu_{3}\nu_{4}} V_{ki}^{*} V_{kj} (1-f_{k}) 
X^{\dagger}_{i,\nu_{3}\nu_{4}} X_{j,\nu_{4}\nu_{2}}   \rho^{\nu_{1}\nu_{3}} c_{i}^{\dagger}  
 \delta(E_{i,\nu_{3}\nu_{4}})  c_{j}  \nonumber\\
&&\hspace{-0.3cm} - \pi \sum_{kij\nu_{3}\nu_{4}} V_{ki}^{*} V_{kj} f_{k} 
X^{\dagger}_{i,\nu_{1}\nu_{3}} X_{j,\nu_{4}\nu_{2}}  c_{i}^{\dagger}  
 \delta(E_{i,\nu_{1}\nu_{3}}) \rho^{\nu_{3}\nu_{4}} c_{j} , \nonumber
\end{eqnarray}
with
\begin{eqnarray}
E_{i,\nu_{a}\nu_{b}} &=& \overline{\epsilon}_{i} + \sum_{j\neq i} \overline{U}_{ij} 
(c^{\dagger}_{j}c_{j}-\delta_{j}) + \Omega(\nu_{a}-\nu_{b}), \nonumber\\
X_{i,\nu_{1}\nu_{2}} &=& \langle \nu_{1} \vert X_{i} \vert \nu_{2} \rangle. \nonumber
\end{eqnarray}
Here, $f_{k}$ denotes the Fermi distribution function of the respective lead, 
L or R, evaluated at energy $\epsilon_{k}$, and $\delta(x)$ stands for the Dirac-delta 
function, where \emph{e.g.}\ 
$\delta(\epsilon+U c_{1}^{\dagger}c_{1})\vert11\rangle=\delta(\epsilon+U)\vert11\rangle$. 
In Eq.\ (\ref{specME}),
we have neglected all principal value terms. 
These terms
describe the renormalization of the molecular 
energy levels due to the molecule-lead coupling \cite{Harbola2006}, 
which 
are negligible for the results discussed below. 
The thus obtained scheme represents a rate equation approach 
\cite{Mitra04,Semmelhack,Siddiqui}.

In the next step, Eq.\ (\ref{specME}) is evaluated with respect to the 
electronic basis functions. Since there is no further approximations involved, 
we present the rather lengthy expressions in appendices \ref{specMEonestate} and 
\ref{specMEtwostates}, where the results for a single electronic state and two 
electronic states can be found, respectively.

\subsection{Observables of interest}
\label{currentandvibex}

\subsubsection{Electronic population and vibrational excitation}

The diagonal elements of the density matrix, $\rho_{a,a}^{\nu\nu}$, encode the 
probability of finding the system in the product state 
$\vert a\rangle\vert\nu\rangle$. 
Hence, for a single electronic state on the molecular bridge 
the occupation of this state is given 
by the expression
\begin{eqnarray}
n_{1} &=& \langle c^{\dagger}_{1} c_{1} \rangle_{H} = \langle c^{\dagger}_{1} c_{1} 
\rangle_{\overline{H}} \\
 &=& \text{tr}_{\text{S}+\text{B}}\lbrace \varrho   c^{\dagger}_{1} c_{1} \rbrace  = 
\text{tr}_{\text{S}}\lbrace \rho  c^{\dagger}_{1} c_{1} \rbrace = 
\sum_{\nu} \rho^{\nu\nu}_{1,1}. \nonumber
\end{eqnarray}
For two electronic states , 
the respective populations are given by
\begin{eqnarray}
n_{1} &=& \langle c^{\dagger}_{1} c_{1} \rangle_{H} = \sum_{\nu} \rho^{\nu\nu}_{11,11} 
+ \rho^{\nu\nu}_{10,10},\\
n_{2} &=& \langle c^{\dagger}_{2} c_{2} \rangle_{H} = \sum_{\nu} \rho^{\nu\nu}_{11,11} 
+ \rho^{\nu\nu}_{01,01}. \nonumber
\end{eqnarray}
Thereby, the subscript $H$/$\overline{H}$ denotes the Hamiltonian, which is 
used to evaluate the respective expectation value. 

The average excitation of the vibrational mode involves a sum over 
all electronic degrees of freedom. For the transport scenario with a single 
electronic state on the molecular bridge, the average vibrational excitation reads
\begin{eqnarray}
\langle a^{\dagger}a \rangle_{H} &=& \langle a^{\dagger} a \rangle_{\overline{H}} + 
\frac{\lambda^{2}}{\Omega^{2}} ( n_{1} - 2 \delta_{1} n_{1} + \delta_{1} ) \\
&=&   \sum_{\nu,a} \nu \rho^{\nu\nu}_{a,a}+ 
\frac{\lambda^{2}}{\Omega^{2}} ( n_{1} - 2 \delta_{1} n_{1} + \delta_{1} ), \nonumber
\end{eqnarray}
and respectively for the transport scenario with two electronic states
\begin{eqnarray}
\langle a^{\dagger}a \rangle_{H} &=&  \sum_{\nu,a} \nu \rho^{\nu\nu}_{a,a} + 
\frac{\lambda_{1}^{2}}{\Omega^{2}} ( n_{1} - 2 \delta_{1} n_{1} + \delta_{1} ) \\
&& + \frac{\lambda_{2}^{2}}{\Omega^{2}} ( n_{2} - 2 \delta_{2} n_{2} + \delta_{2} ) \nonumber\\
&& + 
2\frac{\lambda_{1}\lambda_{2}}{\Omega^{2}} ( \sum_{\nu} \rho^{\nu\nu}_{11,11} -\delta_{2} n_{1} - \delta_{1} n_{2} + \delta_{1}\delta_{2} ) . \nonumber
\end{eqnarray}
Since both operators $c^{\dagger}_{1/2} c_{1/2}$ and $a^{\dagger} a$
act in the subspace of the molecular bridge, the corresponding observables 
are fully determined by the reduced density matrix $\rho$. This is not the case 
for the current operator, which encompasses the bridge 
space and the subspace of the leads.

\subsubsection{Current}

The current through lead $K$, $I_{K}$, is determined by the number of electrons entering or leaving the  
lead per unit time ($K\in\lbrace \text{L,R}\rbrace$)
\begin{eqnarray}
\label{firstcurrent}
I_{K} &=& \langle \hat{I}_{K} \rangle_{H} = -2e \frac{\text{d}}{\text{d} t} 
\sum_{k\in K} \langle c^{\dagger}_{k} c_{k} \rangle_{\overline{H}} \\
&=& 2ie \left[ \sum_{ki} V_{ki} \langle c^{\dagger}_{k} c_{i} X_{i} \rangle_{\overline{H}} 
- \sum_{ki} V_{ki}^{*} \langle c_{i}^{\dagger} X_{i}^{\dagger} c_{k} \rangle_{\overline{H}} 
\right]. \nonumber
\end{eqnarray}
Here, the constant ($-e$) denotes the electron charge and the factor $2$ accounts 
for spin-degeneracy. The specific structure of the current operator requires the 
determination of $Q\varrho=(1-P)\varrho$, since $\text{tr}_{\text{S}+\text{B}}\lbrace 
P\varrho \hat{I}_{K} \rbrace=0$. This projection of the full density matrix has 
already been used to derive Eq.\ (\ref{NakajimaZwanzigEqu}) \cite{Nakajima,Zwanzig},
 and reads
\begin{eqnarray}
\label{qrho}
Q\varrho(t) &=& \text{e}^{-iQ\mathcal{L}t} Q\varrho(0) -i \int_{0}^{t}
\text{d}\tau\,\text{e}^{-i Q\mathcal{L}\tau} Q\mathcal{L} P \,\varrho(t-\tau). 
\nonumber\\
\end{eqnarray}
Using Eq.\ (\ref{qrho}) to evaluate the expression for the current,
Eq.\ (\ref{firstcurrent}), and employing the same approximations that were used for 
the derivation of the master equation, Eq.\ (\ref{genfinalME}), the following
expression for the current through lead $K$ is obtained \cite{May02,Mitra04,Lehmann04,Harbola2006}
\begin{eqnarray}
\label{gencurrentME}
I_{K} &=& -i \int_{0}^{\infty}\text{d}\tau\, \text{tr}_{\text{S}+\text{B}}\lbrace 
\left[ \overline{H}_{\text{SB}}(\tau), \rho_{\text{B}}  \rho \right] \hat{I}_{K} \rbrace .
\end{eqnarray}
In the numerical calculations, Eq.\ (\ref{gencurrentME}) is further evaluated 
within the appropriate basis functions. The explicit formulas are given in appendices 
\ref{specMEonestate} and \ref{specMEtwostates}.
We note that the scheme described above is current conserving, 
\emph{i.e.}\ $I_{\text{L}}=-I_{\text{R}}=I$.

\subsection{Vibration in thermal equilibrium}
\label{thermalequSection}

To analyze and identify vibrational nonequilibrium effects, it is often 
instructive to compare the results of Eqs.\ (\ref{genfinalME}) and (\ref{gencurrentME})
 with those, where the vibrational degree of freedom of the molecular bridge 
is treated as a reservoir's degree of freedom.
To this end, we define a new projection operator $P'$
\begin{eqnarray}
P'\varrho(t)&=&\rho_{\text{B}'} \trRprime{\varrho(t)}\equiv\rho_{\text{B}'}\rho'(t), \\
\rho_{\text{B}'} &=& \mathcal{Z'}^{-1} \text{e}^{-\beta \overline{H}_{\text{B}'}},
\quad \mathcal{Z'}=\trRprime{\text{e}^{-\beta \overline{H}_{\text{B}'}}},
\end{eqnarray}
where 
\begin{eqnarray}
\overline{H}_{\text{B}'}&=& \sum_{k} \epsilon_{k} c_{k}^{\dagger}c_{k} + \Omega a^{\dagger} a.  
\end{eqnarray}
The thus defined reduced density matrix $\rho'$ describes the electronic degrees of
freedom on the molecular bridge.
Analogously to Eq.\ (\ref{genfinalME}) and following the same steps as
outlined  in Section \ref{gentheo},
we obtain the following master equation
\begin{eqnarray}
0 &=& -i \left[ \overline{H}_{\text{S}'} , \rho' \right] \\
&&  - \int_{0}^{\infty} \text{d}\tau\, 
\text{tr}_{\text{B}'}\lbrace \left[ \overline{H}_{\text{S}'\text{B}'} , 
\left[ \overline{H}_{\text{S}'\text{B}'}(\tau), \rho' \rho_{\text{B}'} \right] \right] 
\rbrace,  \nonumber
\end{eqnarray}
with
\begin{eqnarray}
\overline{H}_{\text{S}'}&=& \sum_{i} \overline{\epsilon}_{i} 
c_{i}^{\dagger}c_{i} + \sum_{i<j} \overline{U}_{ij} (c_{i}^{\dagger}c_{i}-
\delta_{i})(c^{\dagger}_{j}c_{j}-\delta_{j}), \nonumber\\ \\
\overline{H}_{\text{S}'\text{B}'} &=& \sum_{ki} 
( V_{ki} X_{i} c_{k}^{\dagger}c_{i} + \text{h.c.} ).
\end{eqnarray} 
Treating the vibrational mode as a reservoir's degree of freedom
does not  exclude the possibility to excite or deexcite it. However, it is assumed that
such a nonequilibrium state relaxes on a short time scale to the thermal equilibrium state.
This can be realized, e.g., by a strong coupling of the vibrational mode 
to a thermal bath.

\section{Results}
\label{secResults}

We have applied the methodology outlined above to various models of
vibrationally coupled electron transport in molecular junctions. The results
described in this Section are structured  according to the complexity of the models
employed and  the associated transport mechanisms.
In Sec.\ \ref{secResonestate} and B we present results for transport through
a molecular bridge with a single and two electronic states, respectively.
Within each of these subsections, we study first symmetric molecular 
junctions,
proceeding with asymmetrically coupled junctions, 
and finally discuss, for transport through two electronic states, the
influence of 
electron-electron interactions. 

In all cases considered the electronic states are coupled to a single
vibrational mode with frequency $\Omega=0.1$\,eV. To represent the vibrational
degree of freedom, the calculations employ
$N_{\text{bas}}=200$  vibrational basis functions, which provide converged
results for all observables and parameters considered in this section.
As is shown in appendix \ref{effectofvibrationalcoherences}, vibrational
coherences have no significant effect and are therefore neglected in these calculations.
The temperature of the leads is set to $10$\,K in all calculations,  since most experiments 
on molecular junctions are carried  out at low temperatures. The same
temperature is used for the vibrational mode in those calculations, where 
the vibration is treated in thermal equilibrium.

\subsection{Transport through a molecular junction with a single electronic state}
\label{secResonestate}

In this section we discuss vibronic effects in transport through a molecular junction 
that involves a single electronic state and a single vibrational mode.
We start with a summary of  the basic mechanisms of resonant emission and absorption 
processes (sketched in Figs.\ \ref{simpleemissionandabsorption} and \ref{el-h-pair-creation}) 
for transport through a state that is symmetrically coupled to the leads.
Furthermore, we discuss  vibrationally induced rectification \cite{Flensberg03b,WuNazinHo04} and 
vibrationally induced negative differential resistance (NDR) 
\cite{Schoeller01,Koch05b,Zazunov06,Dong07,LeRoy,Sapmaz05,Sapmaz06,Leijnse09}
for junctions with  asymmetric coupling to the leads.
Understanding these generic mechanisms facilitates the discussion of
transport through two electronic states in Sec.\ \ref{restwostates} and
extends our previous 
studies \cite{Hartle,Hartle09,Hartle2010}.
In particular, we  show that the  vibrational excitation of a molecular
junction  can  only be understood if 
electron-hole pair creation processes are considered (cf.\ Fig.\ \ref{el-h-pair-creation}) 
\cite{Ueba2002,Ueba2002b,noteonelholepaircreation}. Furthermore, the results demonstrate 
that vibrationally induced rectification is strongly enhanced, if the 
vibrational degree of freedom, due to current-induced local heating, is in a highly 
excited nonequilibrium state.

\subsubsection{Basic mechanisms}
\label{basmech}

We first consider a model with a single electronic state at the molecular
bridge with energy $\epsilon_{1}=0.6$\,eV and a moderate coupling to the
vibrational mode, $\lambda=0.06$\,eV. The left and the right leads are modelled as one-dimensional 
semi-infinite tight-binding chains with a semi-elliptic conduction band, for which 
the respective level-width functions $\Gamma_{\text{L/R}}(E)$ read
\begin{eqnarray}
 \Gamma_{\text{L/R}}(E) &=&  \\
&&\hspace{-1.75cm}\frac{\vert\nu_{\text{L/R}}\vert^{2}}{\vert\beta\vert^{2}} 
\left\{ \begin{array}{ll}
\sqrt{4\vert\beta\vert^{2}-(E-\mu_{\text{L/R}})^{2}}, & 
\vert E-\mu_{\text{L/R}} \vert \leq 2\vert\beta\vert,  \\
0, & \vert E-\mu_{\text{L/R}} \vert > 2\vert\beta\vert. \\
\end{array}
 \right. \nonumber
\end{eqnarray}
Here, $\nu_{\text{L/R}}=0.1$\,eV denote the coupling strengths of the left and the
 right tight-binding chain to the electronic state at the molecular bridge and $\beta=3$\,eV determines 
the band-width in both leads. The difference of the chemical potentials of the leads,
$\mu_{\text{L}}-\mu_{\text{R}}\equiv e\Phi$, determines the applied bias voltage 
$\Phi$, which we assume to drop symmetrically at the contacts, \emph{i.e.}\ 
$\mu_{\text{L}}=-\mu_{\text{R}}=\frac{e}{2}\Phi$.

\begin{figure}
\resizebox{\newwidth}{\newheight}{
\includegraphics{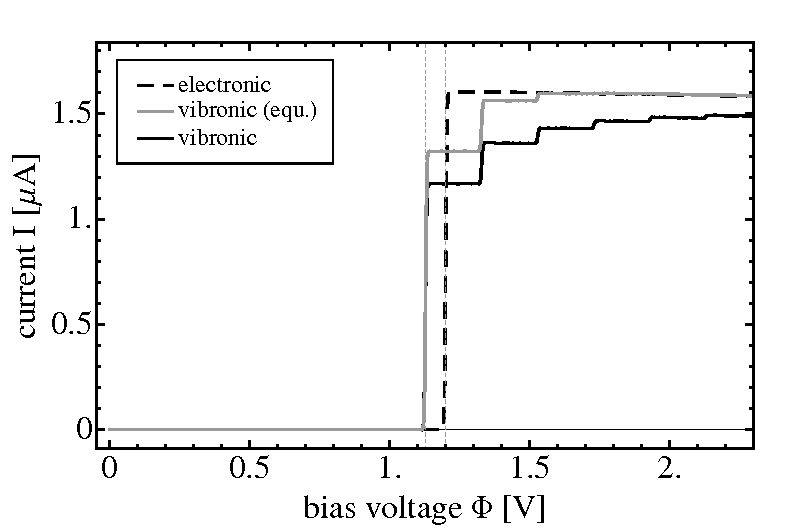}
}
\resizebox{\newwidth}{\newheight}{
\includegraphics{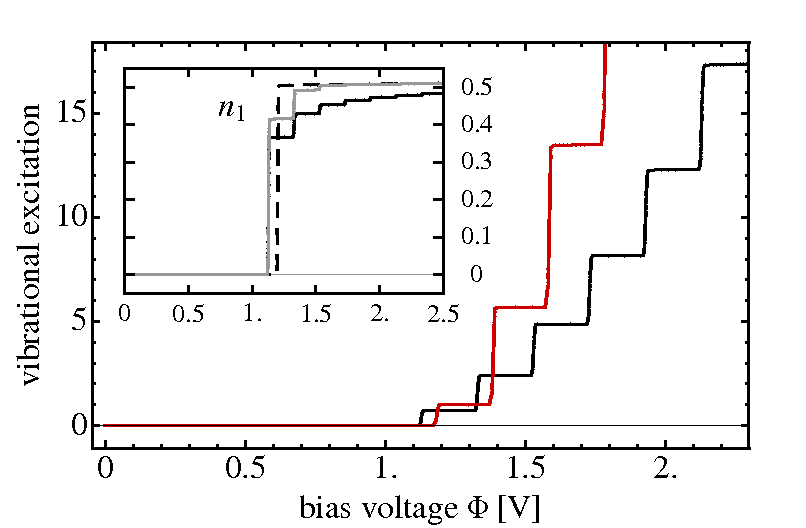}
}
\caption{\label{singlestate} Current and vibrational excitation for a generic model 
system with a single electronic state at the molecular bridge that is symmetrically coupled to the left and 
the right lead, and moderately coupled to a single vibrational mode.  The inset 
shows the corresponding population $n_{1}$ of the electronic level. The dashed 
line refers to a calculation where the electronic-vibrational coupling is set 
to zero. The solid gray and black line are obtained for an electronic-vibrational 
coupling strength $\lambda=0.06$\,eV. Thereby, the gray line is calculated 
employing the thermal equilibrium state of the vibrational mode (at 10K), and 
the black line is obtained with its full current-induced nonequilibrium state. 
The solid red line depicts the vibrational excitation for this model system with 
a reduced electronic-vibrational coupling $\lambda=0.03$\,eV.
}
\end{figure}

Fig.\ \ref{singlestate} shows the corresponding current-voltage characteristics, 
the population of the electronic state at the molecular bridge (inset of Fig.\ \ref{singlestate}b) and the 
vibrational excitation of this system. 
To facilitate the discussion, the dashed line depicts results for a 
purely electronic calculation without vibronic coupling ($\lambda=0$). 
The corresponding current exhibits a single step at the bias voltage $e\Phi=2\epsilon_{1}$, 
indicating that for $e\Phi>2\epsilon_{1}$ electrons from the left lead can resonantly 
tunnel onto the electronic state and further to the right lead. These resonant 
transmission processes result in a current of $I\approx1.6$\,$\mu$A.

\begin{figure}
\begin{tabular}{cccc}
\resizebox{\newwidthprime}{\newheightprime}{
\includegraphics{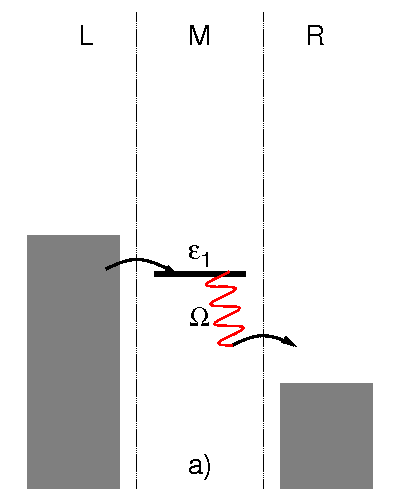}
}
&
\resizebox{\newwidthprime}{\newheightprime}{
\includegraphics{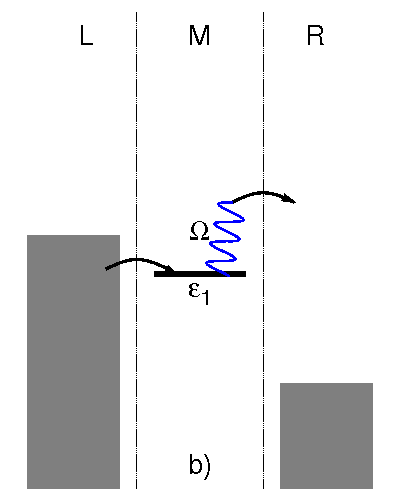}
}
&
\resizebox{\newwidthprime}{\newheightprime}{
\includegraphics{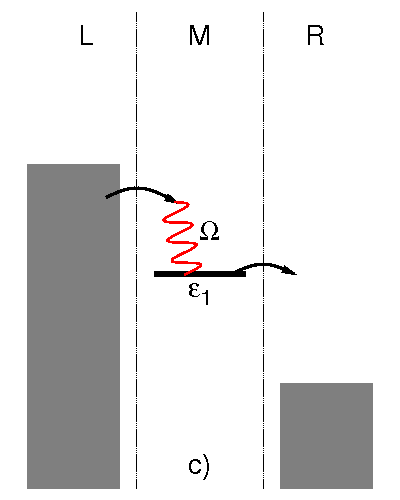}
}
\\ 
\end{tabular}
\caption{\label{simpleemissionandabsorption} Basic schemes of vibrationally 
coupled electron transport processes for a single electronic state. 
Panels a) and c) depict examples for emission processes, where an electron 
sequentially tunnels from the left lead onto the molecule and further to the right lead, 
thereby singly exciting the vibrational mode of the molecular bridge (red wiggly line). 
Such emission processes are effectively 'heating' the junction (local heating). 
An example for a respective absorption process is shown in Panel b), where an 
electron tunnels from the left to the right lead by absorbing a quantum of vibrational 
energy (blue wiggly line). These processes result in local cooling of the junction.}
\end{figure}

\begin{figure}
\begin{tabular}{c}
\resizebox{\newwidthprime}{\newheightprime}{
\includegraphics{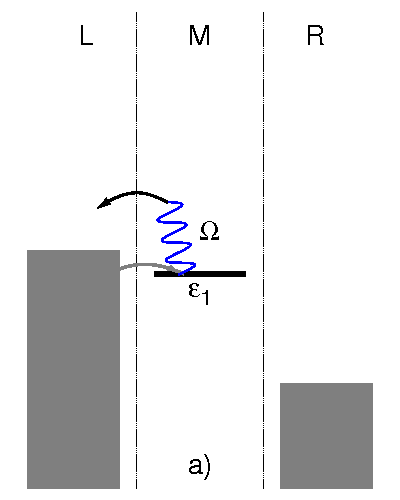}
}
\resizebox{\newwidthprime}{\newheightprime}{
\includegraphics{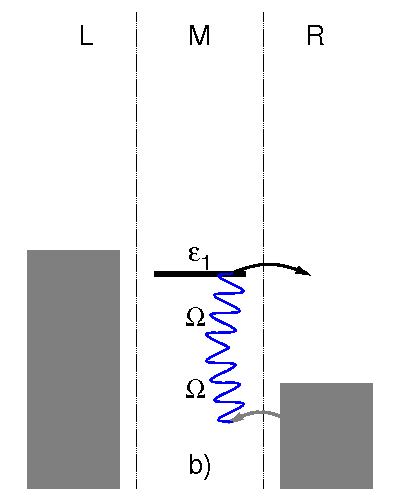}
}
\\ 
\end{tabular}
\caption{\label{el-h-pair-creation} Example processes for electron-hole pair 
creation in a molecular junction. Panel a) depicts an electron-hole pair creation process with respect 
to the left lead by absorption of a single vibrational quantum. 
Panel b) represents an electron-hole pair 
creation process with respect to the right lead by absorption of two vibrational quanta. 
The absorption of two vibrational quanta normally occurs with lower probability.}
\end{figure}

The current-voltage characteristics and the vibrational excitation for a finite vibronic coupling strength of
$\lambda=0.06$\,eV are depicted by the solid black lines.  
For this case, the current rises in a multitude of steps. 
The first step appears at a lower bias voltage $e\Phi=2\overline{\epsilon}_{1}$, 
reflecting the polaron shift of the electronic state. The following steps appear 
at voltages $e\Phi=2\overline{\epsilon}_{1}+2n\Omega$ with $n\in\mathbb{N}$ and 
a step height that gradually decreases with increasing bias voltage. The 
first step at $e\Phi=2\overline{\epsilon}_{1}$ also marks the onset of resonant
transport,  where electrons 
from the left lead can resonantly tunnel onto the electronic state at the
molecular bridge, accompanied by a transition from the initial vibrational
state of the neutral molecule to the same vibrational state of the charged molecule. In a successive 
tunneling process from the molecular bridge to the right lead, which completes the 
transport process, the electronic-vibrational coupling may result in excitation
(Fig.\ \ref{simpleemissionandabsorption}a) or  deexcitation 
(Fig.\ \ref{simpleemissionandabsorption}b) of the vibrational mode. 
At this bias voltage, there are 
$\text{mod}\left(e\Phi,\Omega\right)$  excitation (emission) processes available
(corresponding to the excitation of at most $\text{mod}\left(e\Phi,\Omega\right)$
vibrational quanta, as e.g.\ in Fig.\ \ref{simpleemissionandabsorption}a), 
and a number of deexcitation (absorption) processes 
(Fig.\ \ref{simpleemissionandabsorption}b).
For larger bias voltages, $e\Phi>2(\overline{\epsilon}_{1}+\Omega)$, electrons can 
excite the vibration upon tunneling from the left lead onto the molecule 
(Fig.\ \ref{simpleemissionandabsorption}c), resulting in an increase of current 
and vibrational excitation. The step-wise increase of the current associated with 
these processes gradually becomes smaller. This can be qualitatively rationalized 
by the Franck-Condon (FC) factors 
$\vert X_{0n}\vert^{2}=
\frac{1}{n!}\left(\frac{\lambda}{\Omega}\right)^{2n}\text{e}^{-(\lambda/\Omega)^{2}}$ 
\cite{Zazunov06} that are associated with the transition probability from the vibrational 
ground state to the $n$th excited state. For $\lambda/\Omega<1$, these FC factors decrease with 
increasing $n$. However, a quantitative description of the step-heights is more involved, 
since a variety of emission and absorption processes contribute to 
each step in the current-voltage characteristics \cite{Leon2008}. 

We next consider the current-induced vibrational excitation depicted in
Fig.\ \ref{singlestate}b. 
Similar to the current,
the vibrational excitation increases in a step-wise way with increasing bias
voltage. However, the step
heights  in the vibrational excitation become larger with increasing bias voltage.
This is in striking contrast to the behavior of the current-voltage
characteristics discussed above. 
Analogous to the current, the steps in the vibrational excitation 
are associated with the onset of resonant emission 
processes that involve successively more vibrational quanta. Hence, the relative step 
heights in vibrational excitation are expected to be larger than the relative step heights 
of the respective current-voltage characteristics. However, if only 
vibrational excitation/deexcitation processes induced by electron
transport processes are taken into account, the resulting vibrational
excitation should saturate as does the current. 
Moreover, for a smaller electronic-vibrational coupling a decrease 
of the relative step heights would be expected due to the reduced
Franck-Condon overlap of processes that  involve multiple vibrational quanta.
On the contrary, the comparison of the results for
different vibronic coupling strengths in Fig.\ \ref{singlestate} 
shows the opposite behavior.

A detailed analysis reveals that these intriguing findings are due to vibrationally induced
electron-hole pair creation processes, schematically depicted in Fig.\ \ref{el-h-pair-creation}. 
The process of vibrational relaxation 
due to electron-hole pair creation
is well known from spectroscopic \cite{Eigler1991,Rudge1991,Ho98,Pascual03} and theoretical studies
\cite{Gao1992,Ueba2002,Ueba2002b,noteonelholepaircreation} of 
adsorbates at metal surfaces
but has not been 
analyzed in detail 
in the context of nonequilibrium
transport in molecular
junctions \cite{Schulze08,Romano10,Hartle2010}. 
In molecular junctions, such processes can deexcite the vibrational mode by creation of 
an electron-hole pair in one of the leads
and contribute 
within the same order in the molecule-lead coupling $V_{ki}$ to the
vibrational excitation as transport
induced excitation/deexcitation processes 
(\emph{e.g.}\ the one of Fig.\ \ref{simpleemissionandabsorption}b). 
If the bias voltage is increased, these processes are blocked one by one,
because the  resonant creation of 
an electron-hole pair requires the absorption of increasingly more vibrational quanta.
For typical values of $\lambda$ and the vibrational excitation, this means that the most 
important electron-hole pair creation processes are blocked first. As a
result, the faster increase of the vibrational 
excitation with bias voltage is due to less efficient cooling by electron-hole pair 
creation processes \cite{noteonCoolingbytransportprocesses}. 
Since the blocked electron-hole pair creation processes are even more important for smaller 
electronic-vibrational coupling, the associated step in the vibrational 
excitation characteristics becomes larger for smaller electronic-vibrational coupling. 
As a consequence, for large bias voltages the vibrational mode is more strongly excited the weaker 
it is coupled to the electronic state. This counter-intuitive behavior is
related  to  the phenomenon of vibrational instability \cite{Semmelhack,Hartle09}.
Hence, the observed increase of the step heights in vibrational excitation for 
increasing bias voltages as well as decreasing electronic-vibrational coupling is 
a result of the successive suppression of electron-hole pair creation processes.

The current-induced  vibrational excitation, in turn, has a
pronounced effect on the current-voltage characteristics. This is demonstrated
by the comparison of the solid black line in Fig.\ \ref{singlestate}a with the solid gray line, which is obtained from a 
calculation with the vibration kept in thermal equilibrium (cf.\ Section
\ref{thermalequSection}).
In particular, the result for a thermally equilibrated vibrational 
degree of freedom shows a significantly larger current at a given bias
voltage.
For $2\overline{\epsilon}_{1}<e\Phi<2(\overline{\epsilon}_{1}+\Omega)$, 
where $\langle a^{\dagger}a\rangle\approx0.5$, this behavior can be qualitatively 
understood by considering only the vibrational ground ($\vert\nu=0\rangle$) and 
first excited state ($\vert\nu=1\rangle$). 
For the solid gray line, due to the 
low temperature $T=10K$, the vibrational mode is restricted to its ground-state, 
and the respective transition probability can be characterized by the FC factor 
$\vert X_{00}\vert^{2}$, which describes elastic tunneling processes 
from the left lead onto the molecular bridge. 
Note that at this bias voltage 
all relevant channels for tunneling from the molecule 
to the right lead are open, since $\overline{\epsilon}_{1}-\epsilon_{\text{F}}>5\Omega$.
For the solid black line, the vibrational mode can be 
found in both states, 
and the respective transition probability can be characterized 
by a linear combination of FC factors: 
$(1-\langle a^{\dagger}a\rangle)\vert X_{00}\vert^{2}+\langle a^{\dagger}a\rangle
\left( \vert X_{11}\vert^{2}+\vert X_{01}\vert^{2} \right)$. Thereby, 
the terms $\sim\vert X_{00}\vert^{2}$ and $\sim\vert X_{11}\vert^{2}$ describe 
elastic tunneling processes from the left lead onto the molecular bridge, 
which is either in its vibrational ground- or
first excited state. 
The third term $\sim\vert X_{01}\vert^{2}$ represents inelastic tunneling processes, 
upon which the vibrational mode is deexcited. 
If $\lambda/\Omega<1$, such a linear combination of FC factors
is smaller than $\vert X_{00}\vert^{2}$. 
Hence, for the given model parameters, 
vibrational excitation suppresses the first step in the current.
Our results show that 
this suppression of the current is a rather characteristic phenomenon, 
especially for larger bias voltages.
This can be qualitatively understood with a similar analysis. 
If \emph{e.g.}\ the bias voltage allows for $m=\text{mod}\left(\frac{\Phi}{2}-\overline{\epsilon}_{1},\Omega\right)$ 
resonant emission processes with respect to the left lead 
(cf.\ Fig.\ \ref{simpleemissionandabsorption}c) and if the molecular bridge is in its ground-state (\emph{e.g.}\ at $T=10$\,K), 
the transition probability from the left lead onto the bridge can be given as 
$\sum_{n=0}^{m}\vert X_{0n}\vert^{2}$, which converges to 
unity with increasing $m$ or bias voltage $\Phi$. 
If the vibrational mode, however, is in a nonequilibrium state, where the
population of the $l$th vibrational level is given by $\alpha_{l}\neq\delta_{0l}$ ($l\in\mathbb{N}_{0}$),
the corresponding transition probability is determined by
 $\sum_{l=0..\infty}\alpha_{l}\sum_{n=0}^{l+m}\vert X_{ln}\vert^{2}$.
Because vibrational excitation typically increases much faster than $m$ does with increasing bias voltage, and because
$\sum_{n=0}^{l+m}\vert X_{ln}\vert^{2}\approx1/2$ for $l\gg m$, vibrational excitation typically results in a
lower current.

\subsubsection{Vibrational rectification in asymmetric junctions}
\label{secvibrect}

In many experimental setups, single-molecule junctions are 
non-symmetric with respect to the 
left-right symmetry. This is the case in  STM experiments 
\cite{Ogawa07,Schulze08,Pump08} but often also in break-junction experiments 
\cite{Reichert02,Boehler04,Ballmann2010}. In this and the following section we
study consequences of an asymmetric molecule-lead coupling on
vibrationally coupled electron transport through a single electronic state.
To this end, we employ the same model system as in Sec.\ \ref{basmech} but change 
the coupling of the electronic state to the right lead to $\nu_{\text{R}}=0.01$\,eV.

\begin{figure}
\resizebox{\newwidth}{\newheight}{
\includegraphics{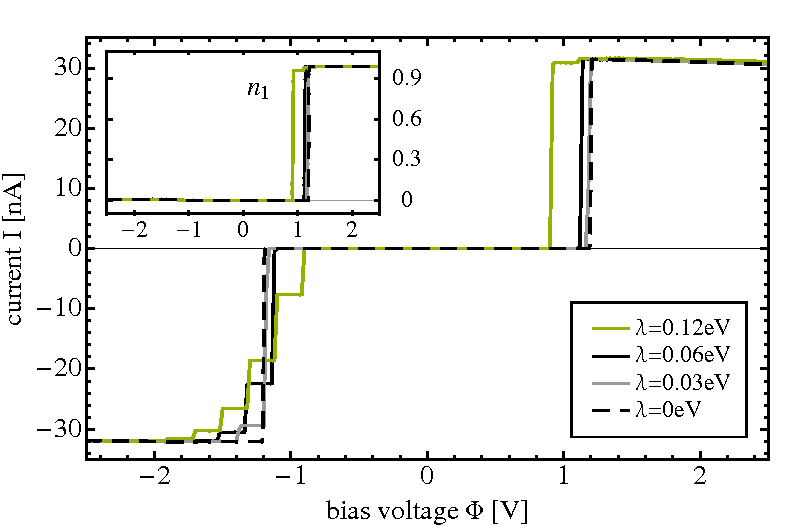}
}
\resizebox{\newwidth}{\newheight}{
\includegraphics{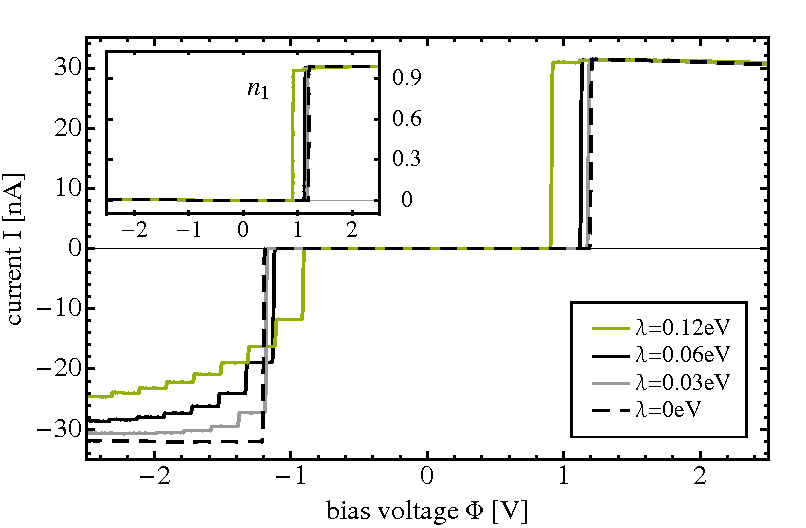}
}
\caption{\label{vibrect} Current-voltage characteristics for a generic model 
system with a single electronic state that is asymmetrically coupled to the leads, and 
moderately coupled to a single vibrational mode with different coupling strengths $\lambda$. 
The current-voltage characteristics of the upper panel have been obtained with the vibrational 
mode kept in thermal equilibrium, while for the ones in the lower panel the
full nonequilibrium state of the vibrational mode is taken into account. 
The insets represent the corresponding populations of the electronic level.
}
\end{figure}

Fig.\ \ref{vibrect} shows the respective current-voltage characteristics for 
different electronic-vibrational couplings $\lambda$. If  the vibration is 
kept in thermal equilibrium (Fig.\ \ref{vibrect}a), the 
corresponding current is  approximately anti-symmetric with respect to bias, 
$I(\Phi) \approx -I(-\Phi)$. 
Significant deviations from this antisymmetry appear around the onset of the 
current at $e\Phi \approx -2\epsilon_{1}$. This can be understood in terms 
of tunneling processes at the boundary between the molecule and the right lead, which represent
the  bottleneck for transport in this asymmetric transport 
scenario with $\nu_{\text{R}}/\nu_{\text{L}}=0.1$. Recall that the probability for
tunneling processes,  where an electron with energy $\overline{\epsilon}_{1}$ enters the bridge 
from the right lead, is $\sim\vert X_{00}\vert^{2}$, 
because the vibrational mode is essentially in its ground-state at $T=10$\,K. 
An electron with energy $\overline{\epsilon}_{1}+\Omega$ entering the 
bridge from the right lead will cause a single excitation of the  
vibrational degree of freedom. Because it is treated as a reservoir degree of
freedom, the vibration will exhibit fast relaxation  
such that the next electron traversing the junction finds the vibrational 
mode again in its ground-state. The respective transition probability is 
 $\sim\vert X_{01}\vert^{2}$. Analogously, for electrons with energy 
$\overline{\epsilon}_{1}+n\Omega$ it is $\sim\vert X_{0n}\vert^{2}$. Thus, for 
negative bias voltages, where electrons flow preferentially from right to
left, 
the current increases with relative step heights that 
are determined by the FC factors $\vert X_{0n}\vert^{2}$. In contrast to the 
findings in Sec.\ \ref{basmech} this is a quantitative statement. 
Thereby, the electronic state remains essentially unoccupied due to the asymmetry 
in the molecule-lead coupling, $\langle c^{\dagger}c\rangle\approx
\frac{\nu_{\text{R}}^{2}}{\nu_{\text{R}}^{2}+\nu_{\text{L}}^{2}}\approx0$ 
(cf.\ the inset of Fig.\ \ref{vibrect}a). For positive biases, on the other
hand,  the molecular electronic state 
becomes almost fully occupied, $\langle c^{\dagger}c\rangle\approx
\frac{\nu_{\text{L}}^{2}}{\nu_{\text{R}}^{2}+\nu_{\text{L}}^{2}}\approx1$, as 
soon as the electronic state enters the bias window. Successive resonant emission 
processes with respect to the left lead are thus Pauli-blocked and do not result 
in further steps in the current-voltage characteristics. 
The step at $e\Phi=2\overline{\epsilon}_{1}$, 
however, is already as high as the one in the electronic current without
vibronic coupling and thus displays 
no suppression due to electronic-vibrational coupling. This is due to the fact 
that a rather large number $\tilde{n}$ of resonant emission processes with respect to the right 
lead is already available at this bias voltage such that 
the sum of the respective FC factors approximately equals unity: 
$\sum_{i=0..\tilde{n}} \vert X_{0i}\vert^{2} \approx 1$. 
This requires the electronic level to be located well above the Fermi energy such that 
$\epsilon_{1}-\epsilon_{\text{F}}\gtrsim\tilde{n}\Omega$. Qualitative similar effects have 
been found previously in theoretical and experimental studies \cite{Flensberg03b,WuNazinHo04}.

Including the current-induced excitation of the vibrational mode in
full nonequilibrium gives qualitatively different results (cf.\ Fig.\ \ref{vibrect}b).
In particular,  for large negative bias voltages the 
current does not approach the maximum value given by the electronic current ($\lambda=0$), 
but remains at a significantly lower absolute value.
Thus, vibronic coupling results 
in a persistent rectification of the current, which appears not only in the vicinity of 
$e\Phi\approx-2\epsilon_{1}$ but for larger absolute values of the bias 
voltage as well. The comparison of Fig.\ \ref{vibrect}a and Fig.\ \ref{vibrect}b shows 
that this vibrational rectification is a pure nonequilibrium effect.

\begin{figure}
\resizebox{\newwidth}{\newheight}{
\includegraphics{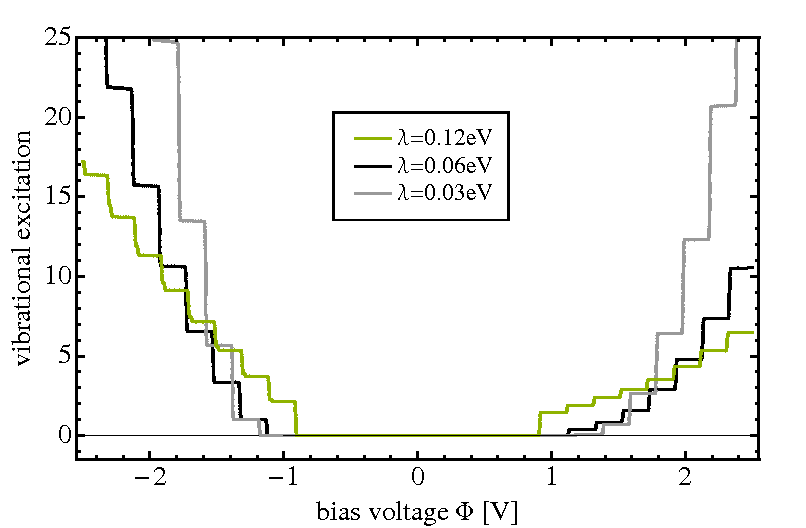}
}
\caption{\label{vibrect2} Average vibrational excitation corresponding to the currents shown 
in Fig.\ \ref{vibrect}b. 
}
\end{figure}

The respective vibrational excitation, depicted in  Fig.\ \ref{vibrect2} 
shows a similar asymmetry as the current. 
For negative bias voltage the level of vibrational excitation is 
much higher than for positive bias. We attribute this behavior to electron-hole pair 
creation processes with respect to the left lead (Fig.\ \ref{el-h-pair-creation}). 
Due to the strong coupling of the left lead to the electronic state at the molecular bridge, these processes provide 
the most important cooling mechanism in the junction. 
For negative bias voltages, however, the creation of an electron-hole pair 
in the left lead requires the absorption of many vibrational quanta ($\geq\text{mod}(2\overline{\epsilon}_{1},\Omega)=12$ in the resonant transport regime),
which is rather unlikely. For positive bias voltages, on the contrary, 
such an electron-hole pair may be generated by absorbing much less vibrational quanta, 
which is much more probable and effectively cooling the vibrational mode.
This results in a lower level of vibrational excitation for positive bias voltages, but also  in a much higher level of vibrational excitation for negative bias voltages. The enhancement of vibrational rectification,
which we observe in Fig.\ \ref{vibrect}, is a result of this higher
level of vibrational excitation
(cf.\ the discussion at the end of Sec.\ \ref{basmech}).

The phenomenon of vibrational rectification 
can be crucial for the 
interpretation of experimental data.  In particular, it explains the 
disappearance or suppression of vibrational side-peaks for a specific direction of the 
bias voltage, although such peaks may be clearly visible for the opposite direction of 
the applied bias voltage. This phenomenon has been observed in a number of
experiments \cite{WuNazinHo04,Schulze08,Ballmann2010}.

\subsubsection{Vibrationally induced Negative Differential Resistance (NDR)}
\label{secvibNDR}

\begin{figure}
\resizebox{\newwidth}{\newheight}{
\includegraphics{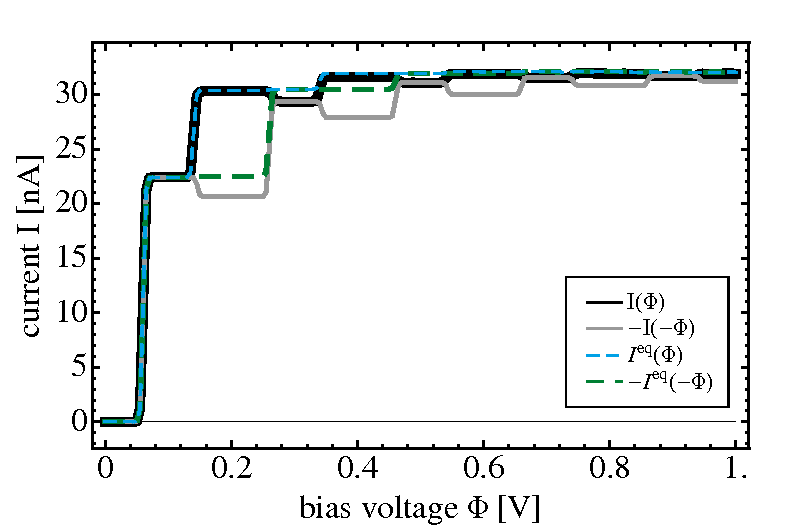}
}
\resizebox{\newwidth}{\newheight}{
\includegraphics{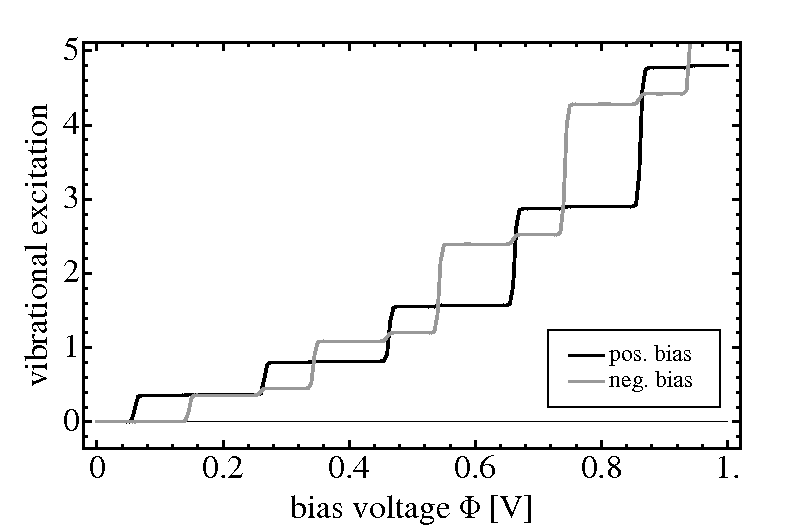}
}
\caption{\label{vibNDR} Current and vibrational excitation for a generic model 
system with a single electronic state close to the Fermi-energy, 
$\vert\overline{\epsilon}_{1}-\epsilon_{\text{F}}\vert<\Omega/2$, and a single vibrational mode. 
The dashed lines refer to a calculation with the vibrational mode in thermal equilibrium, 
while for the solid lines the vibrational mode is treated in its full current-induced 
nonequilibrium state. The current-voltage characteristics for positive (black and blue line) 
and negative bias voltages (gray and green line) are overlayed to highlight areas, 
where $I(\Phi)=-I(-\Phi)$ holds. 
}
\end{figure}

In the model discussed in the previous section, the electronic state was located far 
away from the Fermi level of the system: $\overline{\epsilon}_{1}-\epsilon_{\text{F}}>5\Omega$. 
Thus the first step in the respective current-voltage characteristics involves a 
large number of emission and absorption processes that cannot be resolved separately. 
If the electronic state of such 
a molecular junction is located closer to the Fermi level, \emph{i.e.}\ 
$\vert\overline{\epsilon}_{1}-\epsilon_{\text{F}}\vert<\Omega/2$, every resonant emission 
(or absorption) process can be associated with a corresponding step in the transport 
characteristics. This is illustrated in Fig.\ \ref{vibNDR}, which represents
the transport  characteristics of a model system, where the energy of the
molecular electronic state is $\epsilon_{1}=0.066$\,eV 
($\overline{\epsilon}_{1}=0.03$\,eV). All other parameters are the same as 
in Sec.\ \ref{secvibrect}.

\begin{SCfigure}[3][h]
\resizebox{\newwidthprime}{\newheightprime}{
\includegraphics{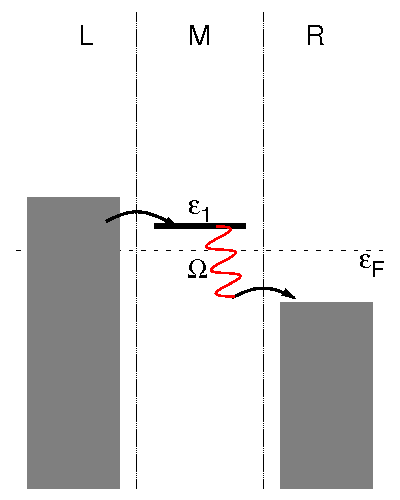}
}
\caption{\label{rightemissionNDR} Example transport processes that become active at a bias voltage $2(\Omega-\overline{\epsilon}_{1})$, as the electronic state of our model system is located close to the Fermi-level of the system, \emph{i.e.}\ $\vert\overline{\epsilon}_{1}-\overline{\epsilon}_{F}\vert<\Omega/2$. If $\overline{\epsilon}_{1}-\overline{\epsilon}_{F}>\Omega/2$, the hopping process from the left lead onto the molecular bridge requires a higher bias voltage, $\Phi=2\overline{\epsilon}_{1}>2(\Omega-\overline{\epsilon}_{1})$, and in that case, these transport processes become active at the same bias voltage as \emph{e.g.}\ elastic transport processes that do not change the vibrational state.}
\end{SCfigure}

We first consider results obtained with keeping the vibrational mode in
thermal equilibrium (at $T=10$\,K, i.e.\ essentially in its ground state), 
depicted by the blue and green lines in Fig.\ \ref{vibNDR}. 
Thereby, the blue line shows the current for 
positive bias voltages, $I(\Phi)$, while the green line represents the current for 
negative bias voltages, $-I(-\Phi)$. Due to the asymmetric molecule-lead coupling, 
$\nu_{\text{R}}/\nu_{\text{L}}=0.1$, distinct resonances appear whenever a 
tunneling process with respect to the right lead becomes available. 
This results, \emph{e.g.}\
for the blue line, in a resonance at $e\Phi=2(\Omega-\overline{\epsilon}_{1})$.
The corresponding transport processes are schematically shown in Fig.\ \ref{rightemissionNDR}.
Similarly, resonances appear at $e\Phi=2\vert n\Omega-\overline{\epsilon}_{1} \vert$, 
with $n\in\mathbb{N}_{0}$, and relative step heights in the current 
that are associated 
with the FC factors $\vert X_{0n}\vert^{2}$ (as already outlined in Sec.\ \ref{secvibrect}). 
Tunneling processes with respect to the right lead appear as steps in 
the green line as well, but at different bias voltages 
$e\Phi=-2(\overline{\epsilon}_{1}+n\Omega)$. 
As a consequence of the different position of the resonances, 
the number of tunneling processes with respect to the 
right lead differs for both lines at specific absolute values of the bias voltage, 
and therefore, 
the blue and the green line encircle 
small rectangles with a width of $4\overline{\epsilon}_{1}$. The heights 
of these rectangles are given by the respective FC factors: 
$\vert X_{0n}\vert^{2}$ ($n\geq1$).
Outside these rectangles, the blue and the green lines have the same absolute values, 
corresponding to an equal number of active transport channels with respect to the right lead.

If the vibrational mode is allowed to develop its full current-induced nonequilibrium 
state, we obtain the solid black line for positive, and the solid gray line for 
negative biases. Again, the black and the gray line encircle rectangles. However, 
the heights of these are enlarged compared to the ones encircled by 
the blue and the green lines.
This is related to the fact that the current represented by the gray line 
decreases at bias voltages $2(\overline{\epsilon}_{1}-n\Omega)$, and is thus significantly
smaller than the one represented by the green line, while the black and the blue line
display essentially the same current values at these bias voltages. 
For $e\Phi=2(n\Omega+\overline{\epsilon}_{1})$ ($n\geq1$), when the black and 
the gray line start to overlap again, we observe a 
significant drop of the current represented by the black line.
Interestingly, for both lines such negative differential resistance 
coincides with the steps in the 
associated vibrational excitation characteristics, which is shown 
in Fig.\ \ref{vibNDR}b.
Here, the black line represents the vibrational excitation induced 
by the current for positive bias voltages, 
while the gray line depicts the one for negative bias voltages. 
For both polarities of the bias voltage, the vibrational excitation
increases at exactly those values of $\Phi$, where electron-hole 
pair creation processes with respect to the left lead become blocked, 
\emph{i.e.} for the black line at $e\Phi=2(n\Omega+\overline{\epsilon}_{1})$ 
and for the gray line at $e\Phi=2(\overline{\epsilon}_{1}-n\Omega)$, 
where $n\geq1$. Recall that 
electron-hole pair creation with respect to the left lead is the most 
important cooling mechanism due to the asymmetry in the molecule-lead coupling, 
$\nu_{\text{L}}=10\nu_{\text{R}}$. Weakening this cooling mechanism results in
larger vibrational excitation, 
and consequently in a reduced current or negative differential resistance
(cf.\ Sec.\ \ref{basmech}).
As a result, 
a vibrational nonequilibrium state may not only induce rectification, as pointed out 
in Sec.\ \ref{secvibrect}, but also negative differential resistance (NDR).

\begin{figure}
\resizebox{\newwidth}{\newheight}{
\includegraphics{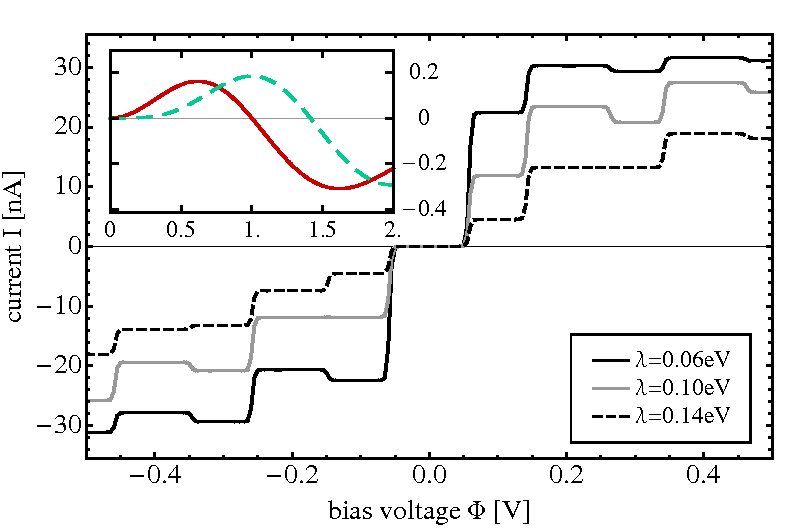}
}
\caption{\label{vibNDRII} Current-voltage characteristics for a single electronic state, 
which is located close to the Fermi-energy,  $\vert\overline{\epsilon}_{1}-\epsilon_{\text{F}}\vert<\Omega/2$, 
and coupled to a single vibrational mode by different coupling strengths $\lambda$. 
Features of NDR disappear one by one for increasing electronic-vibrational coupling. 
The inset shows the difference of the transition probabilities 
$\vert X_{00}\vert^{2}-(\vert X_{11}\vert^{2} + \vert X_{01}\vert^{2})$ (red line) and 
$(\vert X_{00}\vert^{2}+\vert X_{01}\vert^{2})-(\vert X_{11}\vert^{2} + 
\vert X_{01}\vert^{2} + \vert X_{12}\vert^{2})$ (dashed green line) as functions of 
the electronic-vibrational coupling strength $\lambda$. The zero-crossings mark the value of 
the electronic-vibrational coupling, where the respective NDR-feature vanishes.}
\end{figure}

As is shown  in Fig.\ \ref{vibNDRII}, these features of negative 
differential resistance disappear one by one for stronger electronic-vibrational 
coupling $\lambda$. The first one at $e\Phi=2(\overline{\epsilon}_{1}-\Omega)$, 
which is closest to the Fermi level, disappears for 
$\lambda/\Omega\geq1$ \cite{Schoeller01,Zazunov06}. This can be understood, 
as before, by analysis of the tunneling processes with respect to the right lead. 
For bias voltages in the range 
$-2\overline{\epsilon}_{1}>e\Phi>2(\overline{\epsilon}_{1}-\Omega)$ there is only 
a single electronic tunneling process available, which is associated with a 
transition probability $\vert X_{00}\vert^{2}$, because the vibration is not excited 
for this voltage. For smaller bias voltages in the range 
$2(\overline{\epsilon}_{1}-\Omega)>e\Phi>2(\overline{\epsilon}_{1}-2\Omega)$ 
resonant emission processes with respect to the left lead result in a finite 
vibrational excitation. As a consequence, the vibrational mode can be deexcited 
by a resonant absorption process with respect to the right lead. The corresponding 
transition probability for tunneling from the right lead onto the bridge thus 
involves a superposition of $\vert X_{00}\vert^{2}$ and 
$\vert X_{11}\vert^{2}+\vert X_{01}\vert^{2}$. 
For $\lambda/\Omega<1$, such a superposition gives a 
smaller transition probability than the previous one with only $\vert X_{00}\vert^{2}$, 
resulting in an overall smaller current (NDR). For $\lambda/\Omega>1$, this transition 
probability becomes larger than $\vert X_{00}\vert^{2}$. Consequently the 
current is also larger and the NDR at $e\Phi=-2(\Omega-\overline{\epsilon}_{1})$ 
disappears. The red line in the inset of Fig.\ \ref{vibNDRII} shows the difference 
$\vert X_{00}\vert^{2}-(\vert X_{11}\vert^{2} + \vert X_{01}\vert^{2})$ versus the 
electronic-vibrational coupling $\lambda$. The transition from a smaller to a higher 
transition probability is represented by the zero-crossing at $\lambda/\Omega=1$. 
A similar analysis can be done for the second NDR feature at 
$e\Phi=2(\overline{\epsilon}_{1}+\Omega)$, which disappears for $\lambda/\Omega\geq\sqrt{2}$.
This behavior is depicted by the dashed green line in the inset of Fig.\ \ref{vibNDRII}, 
which represents the difference 
$(\vert X_{00}\vert^{2}+\vert X_{01}\vert^{2})-(\vert X_{11}\vert^{2} +
 \vert X_{01}\vert^{2} + \vert X_{12}\vert^{2})$. The NDR features at higher bias voltages 
vanish for increasing electronic-vibrational coupling as well. However, in
this case a detailed analysis 
of all the contributing processes is more involved.

Because this NDR effect is solely based on a vibrational nonequilibrium state, 
it is not exclusively restricted to asymmetric junctions. 
It is also found in 
symmetric junctions (see for example the gray line in Fig.\ 2 of Ref.\ \onlinecite{Hartle09}).
However, in symmetric junctions NDR is usually much less pronounced. 
This can be understood by the transport channels that become active, whenever 
electron-hole pair creation processes become blocked 
(e.g.\ the pair-creation processes in Fig.\ \ref{el-h-pair-creation}a
are blocked as soon as the transport channel represented 
by Fig.\ \ref{simpleemissionandabsorption}c 
becomes active). In contrast to asymmetric junctions, all these transport channels
contribute to the current and thus counteract the NDR mechanism
that is induced by the higher level of vibrational excitation.

\subsection{Transport through a molecular junction with two electronic states}
\label{restwostates}

In this section we study transport through a molecular
junction with two electronic states.  In addition to the phenomena  discussed
above for transport through 
a single electronic state, three new aspects need to be considered \cite{Hartle09}. 
First, in systems with multiple electronic states, the Hamiltonian in Eq.\ (\ref{transformedHamiltonian})
comprises an electron-electron interaction term, which represents both Coulomb interactions and 
vibrationally induced electron-electron correlations.
These electron-electron interactions 
result in a splitting of resonances \cite{Hartle09} that depends on the specific 
population of the electronic states. For asymmetric junctions with a blocking state 
\cite{Hettler2002,Hettler2003,Datta2007}, they can induce strong NDR. While such NDR has been reported 
for specific values of the bias voltage $\Phi$, here we propose another model system 
with a centrally localized electronic state, where NDR due to electronic-vibrational 
coupling extends over a broad range of bias voltages. 
Secondly, coherences of the density matrix can play a significant role in transport 
through multiple electronic states \cite{Esposito09,Esposito2010}. As is shown
in appendix \ref{effectofvibrationalcoherences}, however,  coherences are of
importance only for asymmetrically coupled junctions with quasi-degenerate
energy levels. For the systems considered  in this Section, as in the preceding one,  coherences of 
the density matrix can therefore be neglected.
The third, and most intriguing aspect is that higher-lying electronic states
facilitate the efficient absorption of vibrational energy  \cite{Hartle09,Romano10}.
The role of these resonant absorption processes is elucidated in Sec.\ \ref{ResonantAbsorption} 
for symmetric molecular junctions, and in Sec.\ \ref{ResonantAbsorptionAsym} for 
asymmetric molecular junctions. 
This local cooling mechanism can be very 
efficient and crucial for the stability of the junction. Since closely-lying
electronic states are typical for polyatomic molecules, this mechanism is
expected to be of general importance in molecular junctions.
It is also shown that repulsive Coulomb interactions  may enhance this effect
and thereby significantly improve the  stability of a molecular junction ('Coulomb Cooling'). 
\begin{figure}
\begin{tabular}{c}
\resizebox{\newwidthprime}{\newheightprime}{
\includegraphics{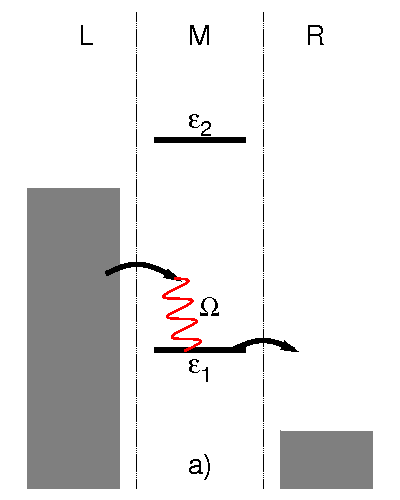}
}
\resizebox{\newwidthprime}{\newheightprime}{
\includegraphics{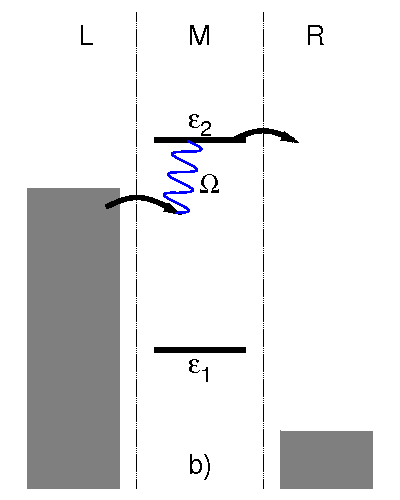}
}
\resizebox{\newwidthprime}{\newheightprime}{
\includegraphics{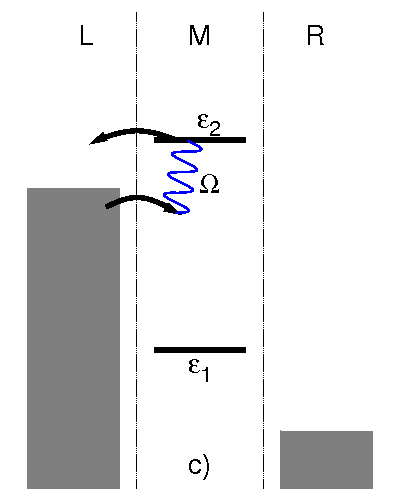}
}
\\ 
\end{tabular}
\caption{\label{twostateprocesses} Basic processes for vibrationally-coupled 
electron transport involving two electronic states, including resonant emission (a), 
resonant absorption (b) and electron-hole pair creation (c) processes.
Due to resonant emission processes with respect to the lower-lying electronic state (panel a)), resonant 
absorption and  electron-hole pair creation processes with respect to the second electronic state 
become active even before this state enters the conduction window set by the applied 
bias voltage.}
\end{figure}

\subsubsection{Resonant absorption processes via a higher-lying electronic 
state in symmetric molecular junctions}
\label{ResonantAbsorption}

We consider a model system with two electronic states, located at 
$\epsilon_{1}=0.15$\,eV and at $\epsilon_{2}=0.8$\,eV above the Fermi-level, 
respectively. The coupling strengths between the vibrational mode and the electronic 
states are given by $\lambda_{1}=0.06$\,eV and $\lambda_{2}=-0.06$\,eV. In the
results discussed in this section no Coulomb 
interactions are taken into account such that electron-electron interactions 
$\overline{U}_{12}=-2\lambda_{1}\lambda_{2}/\Omega$ are induced by vibronic coupling only. 
As before, the left and the right lead are represented by semi-elliptic conduction bands.
The respective level-width functions $\Gamma_{\text{L/R},ij}(E)$ thus read
\begin{eqnarray}
 \Gamma_{\text{L/R},ij}(E) &=&  \\ 
&&\hspace{-2.4cm}
\frac{\nu_{\text{L/R},i}^{*}\nu_{\text{L/R},j}}{\vert\beta\vert^{2}}
\left\{ \begin{array}{ll}
                        \sqrt{4\vert\beta\vert^{2}-(E-\mu_{\text{L/R}})^{2}}, & 
\vert E-\mu_{\text{L/R}} \vert \leq 2\vert\beta\vert,  \\
                        0, & \vert E-\mu_{\text{L/R}} \vert > 2\vert\beta\vert, \\
                       \end{array}
 \right. \nonumber
\end{eqnarray}
where $\nu_{\text{L/R},i}=0.1$\,eV denote the coupling strengths of the left and the 
right lead to the $i$th electronic state and $\beta=3$\,eV determines the band-width 
in both leads.

\begin{figure}
\resizebox{\newwidth}{\newheight}{
\includegraphics{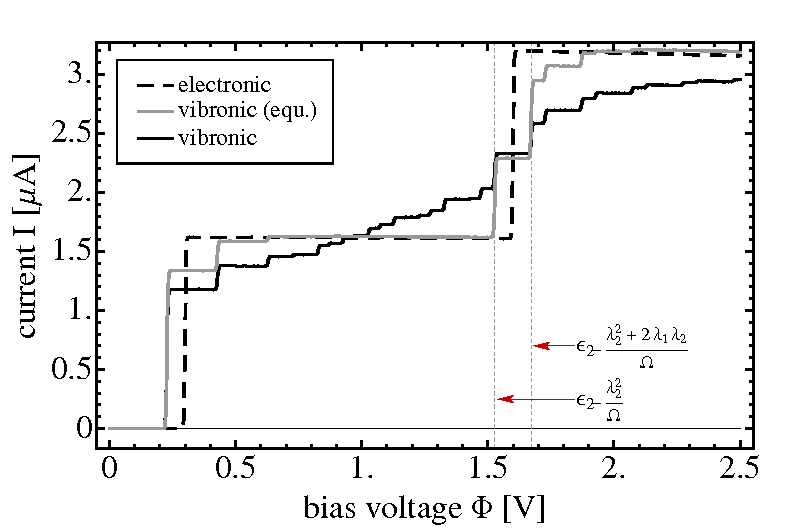}
}
\caption{\label{ResAbPRL} Current-voltage characteristics for a 
model molecular junction comprising two electronic states, which are coupled to a single vibrational mode. 
The dashed line corresponds to a calculation without electronic-vibrational coupling $\lambda_{1/2}=0$. 
The solid black and gray line show results with a moderate coupling of the 
vibrational mode to both electronic states. For the solid gray line the vibrational 
mode is assumed to be in its thermal equilibrium state, while for the solid black line 
the full current-induced nonequilibrium state of the vibrational mode is taken
into account. 
}
\end{figure}

Current-voltage characteristics for this model molecular junction are shown in 
Fig.\ \ref{ResAbPRL}. Thereby, the dashed line represents results without
electronic-vibrational coupling, $\lambda_{1/2}=0$. In this case, the two steps at $e\Phi=2\epsilon_{1}$ 
and $e\Phi=2\epsilon_{2}$ are associated with the onset of transport through the 
lower-lying electronic state at $\epsilon_{1}$ and the higher-lying electronic 
state at $\epsilon_{2}$,  respectively. 

The solid lines depict the results including the coupling of 
the two electronic states to the vibrational mode. We first consider the results (solid gray
line),  where the vibrational mode is kept in thermal equilibrium ($T=10$\,K). 
The respective current-voltage 
characteristics exhibits two major steps 
that correspond to the polaron shifted levels $\overline{\epsilon}_{1/2}$. 
Vibrational side-steps with respect to these resonances can be
distinguished similar to  the results obtained for a single electronic 
state in Sec.\ \ref{basmech} (cf.\ Fig.\ \ref{singlestate}).
A more detailed analysis shows, however, that the current obtained for the two-state
system is not just the sum of the current through the individual states. For
example, the step at 
$e\Phi=2\overline{\epsilon}_{2}$ is just half as high as the one at 
$e\Phi=2\overline{\epsilon}_{1}$. This behavior can be understood in terms of 
vibrationally induced electron-electron interaction, which is not present for 
a single electronic state. In this range of bias voltages, 
$2\overline{\epsilon}_{2}<e\Phi<2(\overline{\epsilon}_{2}-2\lambda_{1}\lambda_{2}/\Omega)$,
electrons in the left lead have not enough energy to doubly occupy 
the molecular bridge. Since the low-lying electronic state is half occupied at 
this bias, $n_{1}=1/2$, the step at $e\Phi=2\overline{\epsilon}_{2}$ is thus reduced 
by a factor of $1-n_{1}=1/2$. Consequently, this step can be associated with the 
electronically excited state of the anion. If the bias exceeds the value 
$2(\overline{\epsilon}_{2}-2\lambda_{1}\lambda_{2}/\Omega)$, electrons from the left 
lead can overcome the additional charging energy $\overline{U}$. Hence, transport 
through the higher-lying electronic state becomes possible even though the low-lying 
electronic state is occupied, \emph{i.e.}\ transport through the di-anionic 
state of the junction. As a result, the  single step at $e\Phi=2\epsilon_{2}$, 
which is associated with the onset of transport through the second electronic state, 
is split into two steps at $e\Phi=2\overline{\epsilon}_{2}$ and 
$e\Phi=2(\overline{\epsilon}_{2}-2\lambda_{1}\lambda_{2}/\Omega)$. Vibrational 
side-steps with respect to the third, di-anionic resonance appear at bias voltages 
$e\Phi=2(\overline{\epsilon}_{2}-2\lambda_{1}\lambda_{2}/\Omega+n\Omega)$ 
($n\in\mathbb{N}$).
Note that $\overline{U}=-2\lambda_{1}\lambda_{2}/\Omega>0$ since the vibrational 
coupling strengths $\lambda_{1}$ and $\lambda_{2}$ differ by sign. 
For $\lambda_{1}\lambda_{2}>0$ the order of the steps associated with the excited 
state of the anion and the di-anionic state would be reversed. 
This scenario is described in Ref.\ \onlinecite{Hartle09}.

\begin{figure}
\resizebox{\newwidth}{\newheight}{
\includegraphics{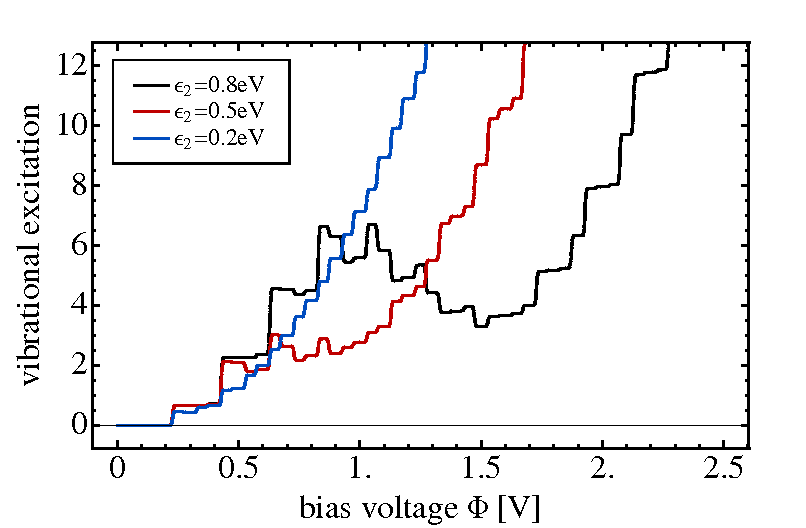}
}
\caption{\label{ResAbPRL2} Vibrational excitation for the two-state model molecular
junction employed for the current-voltage characteristics of Fig.\ \ref{ResAbPRL} (black line). 
The different results correspond to 
different energies $\epsilon_{2}$ for the higher-lying electronic state. 
The solid black line shows the vibrational excitation that corresponds to the 
respective current-voltage characteristics of Fig. \ref{ResAbPRL}. 
The solid blue and red line depict results, where the higher-lying electronic state 
is located closer to the lower-lying electronic level.
}
\end{figure}

Next, we consider the results obtained with the vibrational mode treated in
nonequilibrium (solid black lines of Figs.\ \ref{ResAbPRL} and \ref{ResAbPRL2}). The
current-induced excitation  of the vibrational mode changes the
current-voltage characteristics profoundly. The most striking difference 
is the rise in current even before the second electronic 
state enters the bias window.  This rise of the current is facilitated by the absorption of one 
or more vibrational quanta (see Fig.\ \ref{twostateprocesses}b for an example process). 
These resonant absorption processes appear at voltages 
$e\Phi=2(\overline{\epsilon}_{2}-n\Omega)$ and 
$e\Phi=2(\overline{\epsilon}_{2}-2\lambda_{1}\lambda_{2}/\Omega-n\Omega)$, and can take 
place only if the vibrational mode is in an excited state. Resonant emission processes 
with respect to the lower-lying electronic state (cf.\ Fig.\ \ref{twostateprocesses}a), 
however, are efficiently exciting the vibrational mode, and thus, provide 
the vibrational energy required for these processes (see Fig.\ \ref{ResAbPRL2}). 
In addition, these resonant absorption processes result in a pronounced broadening of the 
resonances that are associated with the second electronic state. This leads to an 
almost Ohmic conductance characteristics, which is observed, e.g.\ in
Fig.\ \ref{ResAbPRL}. This broadening complicates the spectroscopy of
molecular levels in single-molecule junctions \cite{Schoeller01,Stafford09}.

Resonant asorption processes due to higher-lying electronic states 
may have an even more profound effect on the vibrational excitation of the 
molecular junction, as shown in  Fig.\ \ref{ResAbPRL2}.
For lower bias voltages,  $\Phi<0.8$\,V, the vibrational excitation depicted by  the black line  
(which was obtained for the same parameters as the $I$-$\Phi$
curves in Fig.\ \ref{ResAbPRL}) exhibits an increase similar to the case of 
a single electronic state (cf.\ Fig.\ \ref{singlestate}b).  For larger bias voltages, however, 
the vibrational excitation drops by more than 50\%, before it starts 
to increase again for $e\Phi>2\overline{\epsilon}_{2}$. This pronounced
reduction of  vibrational excitation is caused by absorption of vibrational 
energy via resonant absorption processes with respect to the higher-lying
electronic state.  Since the second 
rise of vibrational excitation is shifted by more than $1$\,V, the molecular 
junction is effectively stabilized over a wide range of bias voltages \cite{Hartle09,Romano10}.
It is noted that a similar  decrease of vibrational excitation 
with increasing bias voltage for a molecular junction has been observed in
recent experiments by Ioffe \emph{et al.} \cite{Ioffe08}.

The details of this stabilization or cooling mechanism depend on the energy
gap between the higher- and lower lying 
electronic state. In particular, if the higher-lying state is located too close
to  the lower-lying state (see the solid 
red and blue line in Fig.\ \ref{ResAbPRL2}), resonant emission processes with respect 
to both states become active at the same time, and no decrease 
of vibrational excitation is observed. 
Nevertheless, we expect this cooling mechanism 
to be relevant for most molecular junctions, because polyatomic molecules often exhibit
multiple closely-lying electronic states.

\subsubsection{Resonant absorption processes via a higher-lying electronic state 
in asymmetric molecular junctions}
\label{ResonantAbsorptionAsym}

Resonant absorption processes with respect to a higher-lying electronic state 
involve not only electron transport processes 
(as in Fig.\ \ref{twostateprocesses}b) but also electron-hole pair creation processes 
(sketched in Fig.\ \ref{twostateprocesses}c). 
As was discussed in 
Sec.\ \ref{secResonestate}, electron-hole pair creation plays 
an important role especially for 
asymmetric junctions, where the electronic states on the molecular bridge
 are coupled to the leads with 
different coupling strengths $\nu_{K,i}$. There are eight topologically different 
scenarios for coupling two electronic states asymmetrically to a left and a right lead. 
Since all scenarios show similar effects, we focus in this section 
on the physically most relevant, where the two states of the model system 
introduced in Sec.\ \ref{ResonantAbsorption} are strongly coupled to the left, 
but weakly coupled to the right lead,  $\nu_{\text{L},1/2}=10\nu_{\text{R},1/2}=0.1$\,eV. 
This coupling scenario describes, \emph{e.g.}, the experimental setup of a scanning 
tunneling microscope (STM) \cite{Ho98,Ogawa07,Schulze08,Pump08}.

\begin{figure}
\resizebox{\newwidth}{\newheight}{
\includegraphics{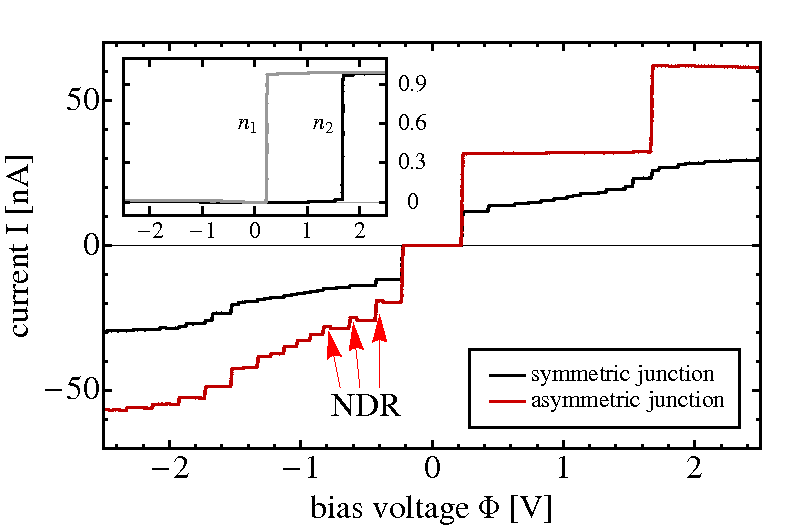}
}
\resizebox{\newwidth}{\newheight}{
\includegraphics{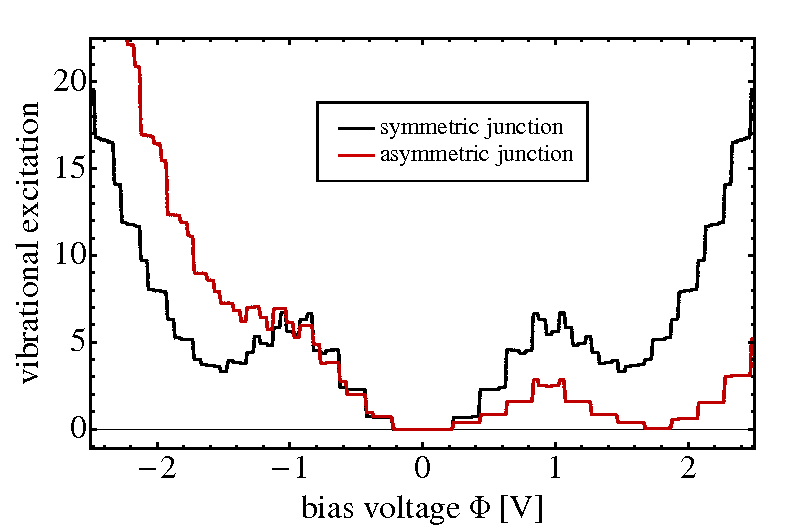}
}
\caption{\label{ResAbAsym} Current-voltage characteristics and vibrational 
excitation for a molecular junction with two electronic states that are moderately 
coupled to a single vibrational mode. The solid red line refers to a calculation with 
asymmetric molecule-lead coupling, where both electronic states are strongly coupled 
to the left lead and weakly coupled to the right lead: $\nu_{\text{L},1/2}=10\nu_{\text{R},1/2}$. 
The solid black lines represent the result for the corresponding symmetric junction, 
which are the same as the solid black lines in Figs.\ \ref{ResAbPRL} and \ref{ResAbPRL2}. The corresponding 
current-voltage characteristic is thereby rescaled by a factor of $1/100$ for a 
better comparison with the red line. The inset in the upper panel shows the population 
of the electronic states $n_{1/2}$ for the asymmetric junction.}
\end{figure}

The corresponding current-voltage characteristics of this model is represented 
by the solid red line in Fig.\ \ref{ResAbAsym}a. For comparison, we also show the 
current of the corresponding symmetric model molecular junction with 
$\nu_{\text{L},1/2}=\nu_{\text{R},1/2}=0.1$\,eV (solid black line) 
rescaled by a factor of $1/100$. For positive bias voltages, the current shows only two steps 
at $e\Phi=2\overline{\epsilon}_{1}$ and 
$e\Phi=2(\overline{\epsilon}_{2}-2\lambda_{1}\lambda_{2}/\Omega)$. There is no 
splitting of the resonances associated with the higher-lying electronic state due 
to electron-electron interactions $\overline{U}$, since the low-lying electronic 
state is almost fully occupied, $n_{1}=1$ (cf.\ the inset of Fig.\ \ref{ResAbAsym}a), 
once it enters the bias window. Transport through the electronically excited state 
of the anion is thus not visible in the respective current-voltage characteristics 
(cf.\ the discussion in Sec.\ \ref{ResonantAbsorption}). For negative biases, 
a multitude of vibronic resonances are seen resulting in the same rectifying 
behavior that was already discussed for a single electronic state in Sec.\ \ref{secvibrect}. 
Small NDR features appear as well (highlighted by red arrows in Fig.\ \ref{ResAbAsym}), 
and can be related to the same mechanisms that was discussed in Sec.\ \ref{secvibNDR}. 
For positive bias voltages resonant absorption via the higher-lying electronic 
state does not significantly contribute to the current, since the vibrational 
mode is efficiently cooled by electron-hole pair creation processes with 
respect to the left lead via 
both electronic states (see, \emph{e.g.}, Fig.\ \ref{twostateprocesses}c). 
These processes are dominant in this regime due to the asymmetry in the 
molecule-lead couplings $\nu_{K,i}$.
For negative biases, however, electron-hole pair creation 
with respect to the left lead
becomes inefficient 
due to the increased amount of vibrational energy 
required for these processes.  
Therefore, vibrational excitation increases faster for negative
bias voltages than for a 
symmetrically coupled molecular junction. 
This asymmetry in vibrational excitation is depicted in Fig.\ \ref{ResAbAsym}b.
Hence, in contrast to our previous findings for a symmetric junction, 
where the steps associated with the higher-lying electronic state 
are significantly broadened, 
we conclude that spectroscopy of molecular orbital energies may be more 
easily performed with a STM-like setup. Although resonant absorption processes are active, 
the mechanism of vibrational rectification restores the signatures of
the individual molecular states.

\subsubsection{Coulomb cooling}
\label{SecCoulombCooling}

In this and the following subsection,  we study the effects of repulsive Coulomb 
interactions ($U>0$) on vibrationally coupled electron transport through 
a molecular junction. We first consider the influence of these interactions 
on resonant absorption processes, and thus on the stability of such a junction.
NDR effects that arise due to repulsive Coulomb interactions 
will be considered in Sec.\ \ref{twostateNDR}.

\begin{figure}
\resizebox{\newwidth}{\newheight}{
\includegraphics{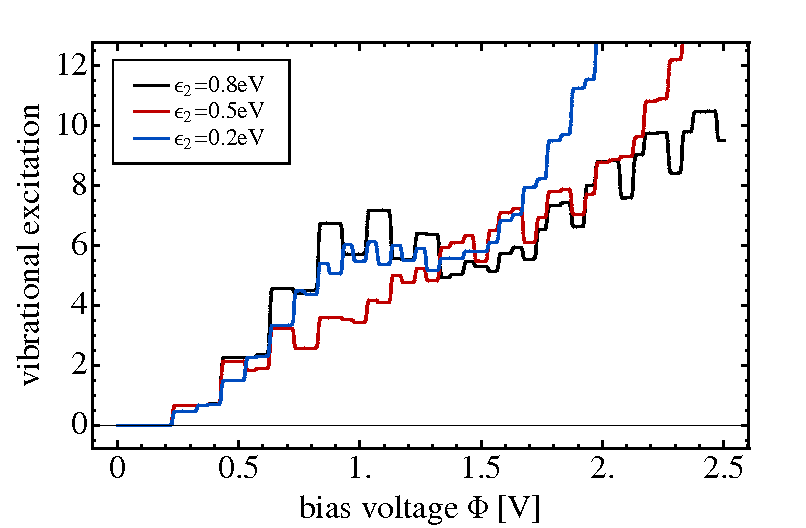}
}
\resizebox{\newwidth}{\newheight}{
\includegraphics{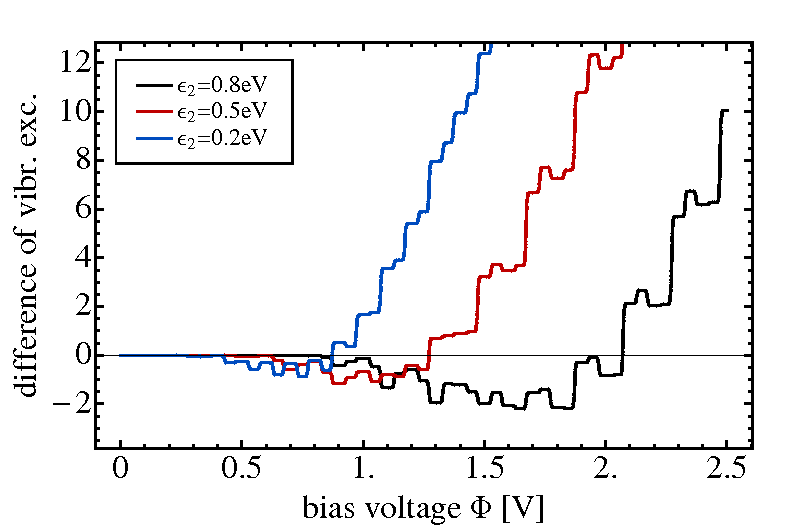}
}
\caption{\label{CoulombCooling} \emph{Upper Panel}: 
Vibrational excitation of a model molecular junction with two electronic states, 
both of which are moderately coupled to a single vibrational mode and symmetrically 
to the leads. In addition, a repulsive electron-electron interaction $U=0.5$\,eV 
is taken into account.
\emph{Lower Panel}: Difference between the vibrational excitation 
shown in Fig.\ \ref{ResAbPRL2} and the upper panel, \emph{i.e.}\  with and without 
Coulomb interaction, 
$\langle a^{\dagger}a\rangle_{H,U=0}-\langle a^{\dagger}a\rangle_{H,U=0.5\text{\,eV}}$.
}
\end{figure}

To this end, we employ the same two-state model system as in Sec.\ \ref{ResonantAbsorption}, 
but add an additional charging energy of $U=0.5$\,eV in the Hamiltonian, Eq.\ (\ref{hel}). 
This term accounts for repulsive Coulomb interactions between two electrons that are occupying 
the two electronic states of the junction. As a consequence, the resonance 
associated with the di-anionic state is shifted towards higher energies, 
from $\overline{\epsilon}_{2}-\frac{2\lambda_{1}\lambda_{2}}{\Omega}$ to 
$\overline{\epsilon}_{2}-\frac{2\lambda_{1}\lambda_{2}}{\Omega}+U$. 
The resulting current-voltage characteristics (data not shown) shows the same 
features as analyzed in Sec.\ \ref{ResonantAbsorption}. The major difference between 
the two scenarios is a shift of all  steps associated with the 
di-anionic state towards higher bias voltages. 
As a result, the current increases more slowly in the resonant transport 
regime for $e\Phi>2\overline{\epsilon}_{1}$. 
The respective vibrational excitation is shown in Fig.\ \ref{CoulombCooling}a. 
Thereby, the solid black, red and blue lines directly correspond to the lines in 
Fig.\ \ref{ResAbPRL2}, which were obtained using the same model parameters but 
without Coulomb interaction.
Since resonant absorption processes with respect to the di-anionic resonance 
are shifted to higher energies, the level of vibrational excitation increases 
faster than without repulsive Coulomb interactions for lower bias voltages. 
At higher bias voltages, however, 
these cooling processes are more efficient than without Coulomb interaction, 
as they require the absorption of less vibrational quanta. Moreover,
they are not competing with absorption processes associated with 
the excited state of the anion. 
As a result, the vibrational excitation, which we obtain for large bias voltages 
including a repulsive Coulomb interaction $U=0.5$\,eV, is  reduced compared 
to that  without additional electron-electron interactions. 
Thus, repulsive  Coulomb interactions effectively stabilize a molecular junction. 
 This cooling mechanism, here referred to as 'Coulomb Cooling'
is analyzed in more detail in Fig.\ \ref{CoulombCooling}b. The three different 
lines depict the difference in vibrational excitation, 
$\langle a^{\dagger}a\rangle_{H,U=0}-\langle a^{\dagger}a\rangle_{H,U=0.5\text{\,eV}}$, 
obtained without, $U=0$\,eV, and with Coulomb interaction, $U=0.5$\,eV, respectively. 
The comparison shows that, except for small bias voltages, the Coulomb interactions lead to
a strong reduction of the vibrational excitation and thus significant cooling \cite{noteonstrongRepulsion}.

\subsubsection{Negative Differential Resistance (NDR) induced by 
Coulomb interactions}
\label{twostateNDR}

In this last section, we study NDR effects induced by Coulomb interaction.

\begin{figure}
\resizebox{\newwidth}{\newheight}{
\includegraphics{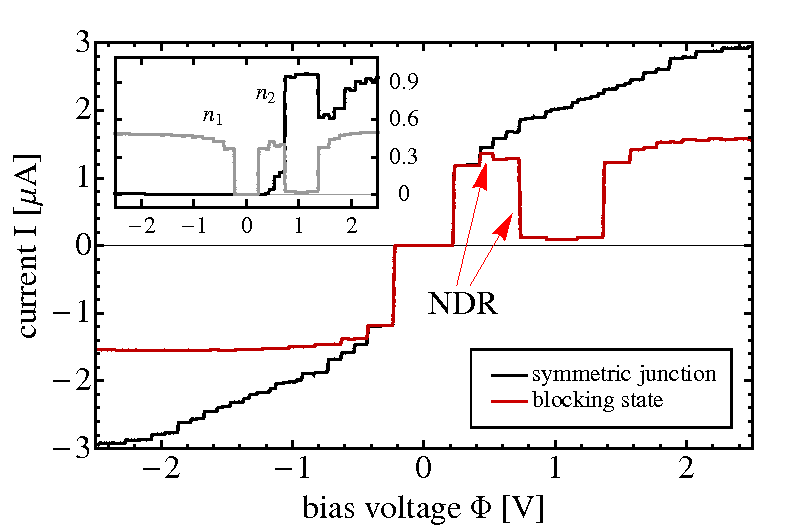}
}
\caption{\label{CoulombAsymm} Current-voltage characteristics for a two-state 
molecular junction with a single vibrational mode, where a repulsive electron-electron 
interaction of $U=0.5$\,eV is taken into account. 
The solid black line represents the current 
for a symmetrically coupled junction, and the solid red line the current for a junction with 
a blocking state \cite{Hettler2002,Hettler2003,Datta2007}, \emph{i.e.}\ a higher-lying state 
that is weakly coupled to the right lead. The populations of the electronic states 
$n_{1/2}$, which correspond to the current-voltage characteristics depicted by the solid red line, 
are shown in the inset. Therein, the solid black line displays the population of the blocking state.  
}
\end{figure}

First, we consider junctions with asymmetric
molecule-lead coupling. As we have already seen, asymmetries in the molecule-lead couplings 
can cause strong changes in the populations of the electronic levels. 
This is particularly important in the presence of Coulomb interactions, where they
may result in pronounced NDR effects. 
To analyze this effect, we employ the same model parameters as in the previous section, except that
the  coupling of the 
higher-lying electronic state to the right lead is reduced to $\nu_{\text{R},2}=0.01$\,eV.
This parameter regime is known as a blocking-state scenario \cite{Hettler2002,Hettler2003,Datta2007}. 
Furthermore, the energy of the higher-lying electronic state 
is chosen as $\epsilon_{2}=0.4$\,eV.
The current-voltage characteristics depicted in  Fig.\ \ref{CoulombAsymm} shows
that the combination of an asymmetry in 
the molecule-lead coupling and Coulomb interactions results in a significant 
NDR effect at a bias voltage of $e\Phi=2\overline{\epsilon}_{2}$. 
Once the bias exceeds this value, $e\Phi>2\overline{\epsilon}_{2}$, 
the second electronic state becomes almost fully occupied (see the inset of 
Fig.\ \ref{CoulombAsymm}). This effectively blocks electron transport through the 
first electronic state, because an additional charging energy $\overline{U}$ is required, and the 
respective current drastically decreases. If the bias is increased further 
$e\Phi>2(\overline{\epsilon}_{1}+\overline{U})$, transport through the first electronic 
state can take place, although the second electronic state remains partially occupied.
For negative bias voltages, however, the second electronic state leaves almost no traces 
in the current-voltage characteristics, as it remains unoccupied. 
As a consequence, NDR is not observed for this direction of the bias voltage.

\begin{figure}
\resizebox{\newwidth}{\newheight}{
\includegraphics{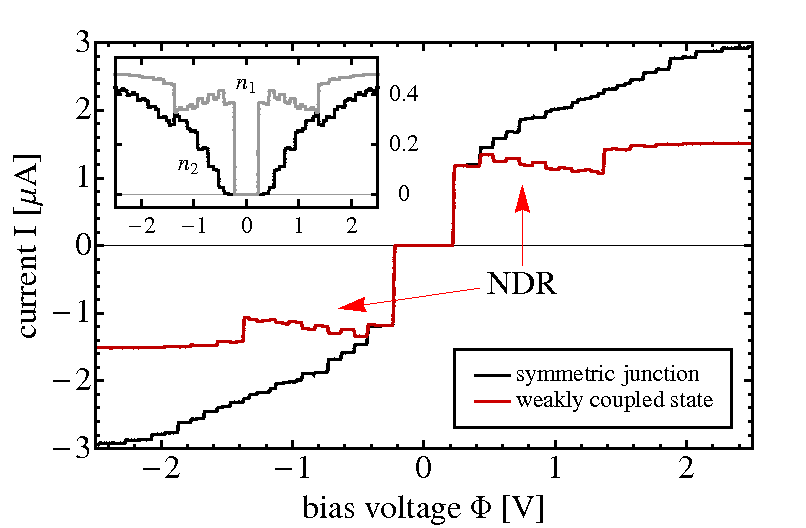}
}
\caption{\label{CoulombAsymm2} Current-voltage characteristics for a symmetrically 
coupled molecular junction, where two electronic states are moderately coupled to a
 single vibrational mode and exhibit repulsive Coulomb interaction, $U=0.5$\,eV. 
The solid black line, as in Fig.\ \ref{CoulombAsymm}, 
represents the result, where both states are coupled to the leads with the 
same coupling strengths $\nu_{K,i}$. For the solid red line the coupling 
of the higher-lying electronic state to both leads was decreased to $\nu_{\text{L/R},2}=0.1\nu_{\text{L/R},1}$. 
}
\end{figure}

Another interesting mechanism for NDR is observed if the higher-lying electronic state of 
the molecular junction is weakly, but symmetrically coupled to both leads, 
$\nu_{\text{L}/\text{r},2}=0.01$\,eV. Such a state corresponds to a molecular orbital, 
which is stronger localized in the central part of the junction. The solid red line in 
Fig.\ \ref{CoulombAsymm2} represents the corresponding current-voltage characteristics. 
In contrast to the NDR effect discussed above, the
current decreases over a broad range of bias voltages, from $2\overline{\epsilon}_{1}$ 
to $2(\overline{\epsilon}_{1}+\overline{U})$. Even for negative bias voltages, 
the absolute value of the current decreases from $-2\overline{\epsilon}_{1}$ to 
$-2(\overline{\epsilon}_{1}+\overline{U})$. Hence, this mechanism for NDR is 
symmetric with respect to bias. The decrease in the current is accompanied by an 
increase in the population of the second electronic state, which, as before, is 
blocking the current through the first electronic state. The inset of 
Fig.\ \ref{CoulombAsymm2} depicts the populations $n_{1/2}$ of both electronic states
and shows the successive increase of $n_{2}$. Thereby, the higher-lying state, 
due to resonant absorption processes, becomes 
occupied even before it enters the bias window. 
Increasing the bias voltage, the occupation number $n_{2}$ increases even further, 
as electronic tunneling processes and resonant emission processes become active. 
Overall, this increase takes place in small steps, resulting in
 an almost smooth current-voltage characteristics. 
Thus, transport through the lower-lying electronic state is not 
blocked in a single step, as for the NDR discussed before, but in a succession of several
small steps, which results  in a NDR feature that extends over a broad range of bias voltages.

\section{Conclusion}

In this paper we have studied vibrationally coupled electron transport in 
single-molecule junctions. 
The study was based on generic models for
molecular junctions, which include electronic states on the molecular
bridge that are
vibrationally coupled and exhibit Coulomb interaction. The transport
calculations employed a master equation approach, which is based 
on a second order expansion in the molecule-lead coupling. 
To the given order in the molecule-lead coupling, electronic-vibrational coupling 
and Hubbard-like electron-electron interactions are accurately described in
this approach. 

The results obtained for a series of models with increasing complexity, show a
multitude of interesting transport phenomena, including vibrational
excitation, rectification, negative differential resistance (NDR) as well as
local cooling. While some of these phenomena have been observed or
proposed before, the present analysis extends previous studies and allows a
more detailed understanding of the underlying transport mechanisms.
In particular, the analysis shows that many of the observed phenomena cannot be  explained if only
transport-induced processes  (cf.\ Fig.\ \ref{simpleemissionandabsorption}) are
taken into account, but also require the consideration of electron-hole pair 
creation processes (cf.\ Fig.\ \ref{el-h-pair-creation}).
For example, these processes explain the increase of the step heights in 
vibrational excitation with increasing bias voltage and/or 
smaller electronic-vibrational coupling. 
Our results also show that in junctions with asymmetric molecule-lead coupling, electronic-vibrational
interaction may result in a significant 
rectification of the current and the vibrational excitation, 
and even negative differential resistance.
Both phenomena are accompanied by an highly excited nonequilibrium state 
of the vibrational mode. In these cases, electron-hole pair creation processes 
play a key-role. In contrast to transport-induced processes, 
they involve only one of the leads, and thus, effectively transfer the asymmetry 
of the molecule-lead coupling to the transport characteristics. 

Extending previous work \cite{Hartle09}, we have also given a detailed analysis of the influence of 
multiple electronic states on vibronic effects in molecular junctions.
The results show that resonant absorption processes involving
higher-lying electronic states efficiently reduce the level of
vibrational excitation and thus can stabilize a molecular junction over a broad range of 
bias voltages \cite{Hartle09,Romano10}.
In this context, we have shown  that 
repulsive Coulomb interactions, which shift these resonant 
absorption processes to higher energies, can strongly enhance this
cooling mechanism ('Coulomb Cooling') and thus improve the stability of the junction. 
Repulsive Coulomb interaction may also cause pronounced negative differential
resistance (NDR).  While such
NDR effects were reported before at specific values of the bias voltage \cite{Hettler2002,Hettler2003,Datta2007}, 
in the model proposed in this work NDR  is caused by vibronic coupling and
extents over a broad range of bias voltages.
It should be noted that single-molecule junctions often exhibit a number of closely-lying electronic states, 
which are strongly coupled with the junctions vibrational degrees of freedom. 
Thus, it can be expected that the phenomena analyzed in this work are of 
relevance for many molecular junctions and may facilitate the understanding and 
interpretation of experimental data.

\emph{Acknowledgment:}\ We thank M.\ \v{C}\'{\i}\v{z}ek, I. Pshenichnyuk, 
R. Volkovich and U.\ Peskin for helpful 
and inspiring discussions. 
The generous allocation of computing time by the Leibniz 
Rechenzentrum M\"unchen (LRZ) as well as financial support from  
the Deutsche Forschungsgemeinschaft (DFG), the European Cooperation 
in Science and Technology (COST), the German-Israeli Foundation for Scientific 
Development (GIF), and the Fonds der Chemischen Industrie (FCI)  are gratefully acknowledged.
This work was carried
out in the framework of the Cluster of Excellence 'Engineering
of Advanced Materials' of the DFG.
R.\ H.\ gratefully acknowledges the hospitality in the group of M.\ \v{C}\'{\i}\v{z}ek 
in Prague and in the group of U.\ Peskin in Haifa.

\appendix

\section{Master equation and current formula for a single electronic state}
\label{specMEonestate}

In this appendix, we give the explicit formulas for the master equation, Eq.\ (\ref{specME}), 
and the current, Eq.\ (\ref{gencurrentME}), for transport through a single electronic 
state coupled to a single vibrational mode. The density matrix of this scenario consists 
of $4N_{\text{bas}}^{2}$ elements, $\rho_{0,0}^{\nu_{1}\nu_{2}},\rho_{1,0}^{\nu_{1}\nu_{2}},
\rho_{0,1}^{\nu_{1}\nu_{2}},\rho_{1,1}^{\nu_{1}\nu_{2}}$, corresponding to the two charge 
states $\vert0\rangle$ and $\vert1\rangle$. In Eqs.\ (\ref{specME}) and (\ref{gencurrentME}) 
coherences between different charge states $\rho_{0,1}^{\nu_{1}\nu_{2}}$ and 
$\rho_{1,0}^{\nu_{1}\nu_{2}}$ are not coupled with the elements $\rho_{0,0}^{\nu_{1}\nu_{2}}$ 
and $\rho_{1,1}^{\nu_{1}\nu_{2}}$ that are diagonal with respect to the electronic 
degrees of freedom. Therefore, these elements do not need to be considered in the following. 
If we evaluate  Eq.\ (\ref{specME}) in between the charge states $\langle0\vert$ 
and $\vert 0 \rangle$, we obtain
\begin{eqnarray}
\label{singleDMT}
\label{singleME}
&& - 2 i \Omega (\nu_{1}-\nu_{2}) \rho^{\nu_{1}\nu_{2}}_{0,0} = \\
&&\phantom{+} \sum_{\nu_{3}\nu_{4}K}  f_{K}(E_{\nu_{3}\nu_{4}}) 
\Gamma_{K}(E_{\nu_{3}\nu_{4}}) X_{\nu_{1}\nu_{3}} X_{\nu_{3}\nu_{4}}^{\dagger} 
\rho_{0,0}^{\nu_{4}\nu_{2}} \nonumber\\
&&  -  \sum_{\nu_{3}\nu_{4}K}  (1-f_{K}(E_{\nu_{4}\nu_{2}}) ) 
\Gamma_{K}(E_{\nu_{4}\nu_{2}}) X_{\nu_{1}\nu_{3}} X_{\nu_{4}\nu_{2}}^{\dagger}
 \rho_{1,1}^{\nu_{3}\nu_{4}}  \nonumber\\
&& -  \sum_{\nu_{3}\nu_{4}K}  (1-f_{K}(E_{\nu_{3}\nu_{1}}) ) 
\Gamma_{K}(E_{\nu_{3}\nu_{1}}) X_{\nu_{1}\nu_{3}} X_{\nu_{4}\nu_{2}}^{\dagger} 
\rho_{1,1}^{\nu_{3}\nu_{4}} \nonumber\\
&& + \sum_{\nu_{3}\nu_{4}K}  f_{K}(E_{\nu_{4}\nu_{3}}) 
\Gamma_{K}(E_{\nu_{4}\nu_{3}}) X_{\nu_{3}\nu_{4}} X_{\nu_{4}\nu_{2}}^{\dagger} 
\rho_{0,0}^{\nu_{1}\nu_{3}}, \nonumber
\end{eqnarray}
with
\begin{eqnarray}
&&E_{\nu_{a}\nu_{b}} = \overline{\epsilon}_{1} + \Omega(\nu_{a}-\nu_{b}). \nonumber
\end{eqnarray}
The corresponding $\langle1\vert..\vert1\rangle$ component reads
\begin{eqnarray}
\label{singleDMT2}
\label{singleME2}
&& - 2 i \Omega (\nu_{1}-\nu_{2}) \rho^{\nu_{1}\nu_{2}}_{1,1} = \\
&&\phantom{+}\sum_{\nu_{3}\nu_{4}K} (1-f_{K}(E_{\nu_{4}\nu_{3}}))
 \Gamma_{K}(E_{\nu_{4}\nu_{3}}) X_{\nu_{3}\nu_{4}} X_{\nu_{1}\nu_{3}}^{\dagger} 
\rho_{1,1}^{\nu_{4}\nu_{2}} \nonumber\\
&& -  \sum_{\nu_{3}\nu_{4}K}  f_{K}(E_{\nu_{2}\nu_{4}})
 \Gamma_{K}(E_{\nu_{2}\nu_{4}}) X_{\nu_{4}\nu_{2}}  X_{\nu_{1}\nu_{3}}^{\dagger}
\rho_{0,0}^{\nu_{3}\nu_{4}}  \nonumber\\
&& -  \sum_{\nu_{3}\nu_{4}K}  f_{K}(E_{\nu_{1}\nu_{3}}) 
\Gamma_{K}(E_{\nu_{1}\nu_{3}}) X_{\nu_{4}\nu_{2}} X_{\nu_{1}\nu_{3}}^{\dagger} 
\rho_{0,0}^{\nu_{3}\nu_{4}} \nonumber\\
&& +  \sum_{\nu_{3}\nu_{4}K}  (1-f_{K}(E_{\nu_{3}\nu_{4}})) 
\Gamma_{K}(E_{\nu_{3}\nu_{4}}) X_{\nu_{4}\nu_{2}} X_{\nu_{3}\nu_{4}}^{\dagger}
\rho_{1,1}^{\nu_{1}\nu_{3}}.  \nonumber
\end{eqnarray}
From this set of algebraic equations, Eqs.\ (\ref{singleME}) and (\ref{singleME2}), 
we determine the elements of the reduced density matrix, which are used to compute 
the current-voltage characteristics and the corresponding vibrational excitation 
of a single-molecule junction represented by a single electronic state (cf.\ Sec.\,\ref{currentandvibex}). 
Thereby, the current through a single electronic state is explicitly given by
\begin{eqnarray}
\label{singlecurrent}
 I_{K} &=& \\ 
&&\hspace{-1.cm} -e \sum_{\nu_{1}\nu_{2}\nu_{3}} 
 \left(1-f_{K}(E_{\nu_{2}\nu_{1}})\right) \Gamma_{K}(E_{\nu_{2}\nu_{1}})  X_{\nu_{1}\nu_{2}}(\tau)
  X_{\nu_{3}\nu_{1}}^{\dagger} \rho_{1,1}^{\nu_{2}\nu_{3}} \nonumber\\
&&\hspace{-1.cm} + e \sum_{\nu_{1}\nu_{2}\nu_{3}} 
 f_{K}(E_{\nu_{1}\nu_{2}})  \Gamma_{K}(E_{\nu_{1}\nu_{2}}) X_{\nu_{3}\nu_{1}}
 X^{\dagger}_{\nu_{1}\nu_{2}}
   \rho_{0,0}^{\nu_{2}\nu_{3}} \nonumber\\ 
&&\hspace{-1.cm} + e \sum_{\nu_{1}\nu_{2}\nu_{3}} 
 f_{K}(E_{\nu_{3}\nu_{2}})  \Gamma_{K}(E_{\nu_{3}\nu_{2}})
  X_{\nu_{2}\nu_{3}} X_{\nu_{3}\nu_{1}}^{\dagger} \rho_{0,0}^{\nu_{1}\nu_{2}}  \nonumber\\ 
&&\hspace{-1.cm} - e \sum_{\nu_{1}\nu_{2}\nu_{3}} 
 \left(1-f_{K}(E_{\nu_{2}\nu_{3}})\right) \Gamma_{K}(E_{\nu_{2}\nu_{3}})
    X_{\nu_{3}\nu_{1}} X_{\nu_{2}\nu_{3}}^{\dagger} \rho^{\nu_{1}\nu_{2}}_{1,1}. \nonumber
\end{eqnarray}
In Eq.\ (\ref{singlecurrent}), principal value terms, 
as for the computation of the reduced density matrix, are disregarded.

\section{Master equation and current formula for two electronic states}
\label{specMEtwostates}

In this appendix, we give the explicit formulas for the master equation, Eq.\ (\ref{specME}), 
and the current, Eq.\ (\ref{gencurrentME}), for transport through two electronic states 
coupled to a single vibrational mode. Similar to the scenario with one electronic state, 
coherences between different charge states are not considered, as they are 
not coupled with the remaining elements of the density matrix. Thus, we are left with 
the computation of six different density matrix elements: $\rho_{00,00}^{\nu_{1}\nu_{2}}$, 
$\rho_{i,j}^{\nu_{1}\nu_{2}}$ ($i,j\in\lbrace1,2\rbrace$), and $\rho_{11,11}^{\nu_{1}\nu_{2}}$, 
where we employ the states $\vert00\rangle$, $\vert1\rangle\equiv\vert10\rangle$, 
$\vert2\rangle\equiv\vert01\rangle$, and $\vert11\rangle$. For notational convenience 
we introduce indices $\overline{i}$, defined by $\overline{1}=2$ and $\overline{2}=1$. \\
Accordingly, we evaluate Eq.\ (\ref{specME}) between the states 
$\langle 00 \vert$ and $\vert 00 \rangle$
\begin{eqnarray}
\label{twostatesME0000}
- 2 i \Omega(\nu_{1}-\nu_{2}) \rho_{00,00}^{\nu_{1}\nu_{2}} &=& \\
&&\hspace{-3.75cm} \phantom{+} \sum_{i\nu_{3}\nu_{4}K}
 \Gamma_{K,ii}(E^{-}_{i\nu_{3}\nu_{4}})
 f_{K}(E^{-}_{i\nu_{3}\nu_{4}}) X_{i,\nu_{1}\nu_{3}} X_{i,\nu_{3}\nu_{4}}^{\dagger} 
  \rho_{00,00}^{\nu_{4}\nu_{2}}  \nonumber\\
&&\hspace{-3.75cm} - \sum_{ij\nu_{3}\nu_{4}K}
 \Gamma_{K,ji}(E^{-}_{j\nu_{4}\nu_{2}}) 
(1-f_{K}(E^{-}_{j\nu_{4}\nu_{2}})) X_{i,\nu_{1}\nu_{3}} X_{j,\nu_{4}\nu_{2}}^{\dagger} 
 \rho_{i,j}^{\nu_{3}\nu_{4}} \nonumber\\
&&\hspace{-3.75cm} - \sum_{ij\nu_{3}\nu_{4}K}
 \Gamma_{K,ji}(E^{-}_{i\nu_{3}\nu_{1}})
 (1-f_{K}(E^{-}_{i\nu_{3}\nu_{1}})) X_{i,\nu_{1}\nu_{3}} X_{j,\nu_{4}\nu_{2}}^{\dagger} 
 \rho_{i,j}^{\nu_{3}\nu_{4}}  \nonumber\\
&&\hspace{-3.75cm} + \sum_{i\nu_{3}\nu_{4}K}
 \Gamma_{K,ii}(E^{-}_{i\nu_{4}\nu_{3}})
 f_{K}(E^{-}_{i\nu_{4}\nu_{3}}) X_{i,\nu_{3}\nu_{4}} X_{i,\nu_{4}\nu_{2}}^{\dagger}  
\rho^{\nu_{1}\nu_{3}}_{00,00},  \nonumber
\end{eqnarray}
with
\begin{eqnarray}
E_{i\nu_{a}\nu_{b}}^{-} &=& \overline{\epsilon}_{i} - 
\overline{U}_{12} \delta_{\overline{i}} + \Omega(\nu_{a}-\nu_{b}). \nonumber
\end{eqnarray}
Evaluation of Eq.\ (\ref{specME}) between the states 
$\langle a \vert$ and $\vert b \rangle$, where 
$a,b\in\lbrace \vert10\rangle,\vert01\rangle\rbrace$, gives
\begin{eqnarray}
\label{twostatesME0000II}
- 2 i \left( \Omega (\nu_{1}-\nu_{2} ) + \epsilon_{a} - \epsilon_{b} \right) \rho^{\nu_{1}\nu_{2}}_{a,b} &=& \\ 
&&\hspace{-5.5cm}\phantom{+} \sum_{i\nu_{3}\nu_{4}K} 
\Gamma_{K,i\overline{a}}(E_{i,\nu_{3}\nu_{4}}^{+}) f_{K}(E_{i,\nu_{3}\nu_{4}}^{+}) 
X_{\overline{a},\nu_{1}\nu_{3}} X_{i,\nu_{3}\nu_{4}}^{\dagger} 
\rho^{\nu_{4}\nu_{2}}_{\overline{i},b} (-1)^{a+\overline{i}} \nonumber\\
&&\hspace{-5.5cm} + \sum_{j\nu_{3}\nu_{4}K} 
\Gamma_{K,aj}(E^{-}_{j,\nu_{4}\nu_{3}}) (1-f_{K}(E^{-}_{j,\nu_{4}\nu_{3}})) 
 X_{j,\nu_{3}\nu_{4}} X^{\dagger}_{a,\nu_{1}\nu_{3}} \rho^{\nu_{4}\nu_{2}}_{j,b} 
 \nonumber\\
&&\hspace{-5.5cm} - \sum_{\nu_{3}\nu_{4}K}
 \Gamma_{K,\overline{b}\overline{a}}(E_{\overline{b},\nu_{4}\nu_{2}}^{+}) 
(1-f_{K}(E_{\overline{b},\nu_{4}\nu_{2}}^{+})) X_{\overline{a},\nu_{1}\nu_{3}}
 X_{\overline{b},\nu_{4}\nu_{2}}^{\dagger} \nonumber\\
&&\hspace{1.2cm}  \cdot \rho^{\nu_{3}\nu_{4}}_{11,11}  (-1)^{a+b}  \nonumber\\
&&\hspace{-5.5cm} - \sum_{\nu_{3}\nu_{4}K}
 \Gamma_{K,ab}(E^{-}_{b,\nu_{2}\nu_{4}}) f_{K}(E^{-}_{b,\nu_{2}\nu_{4}}) 
X_{b,\nu_{4}\nu_{2}} X^{\dagger}_{a,\nu_{1}\nu_{3}}  \rho^{\nu_{3}\nu_{4}}_{00,00}
    \nonumber\\
&&\hspace{-5.5cm} - \sum_{\nu_{3}\nu_{4}K} 
\Gamma_{K,\overline{b}\overline{a}}(E_{\overline{a},\nu_{3}\nu_{1}}^{+}) 
(1-f_{K}(E_{\overline{a},\nu_{3}\nu_{1}}^{+})) X_{\overline{a},\nu_{1}\nu_{3}}
 X_{\overline{b},\nu_{4}\nu_{2}}^{\dagger} \nonumber\\
&&\hspace{1.2cm} \cdot  \rho^{\nu_{3}\nu_{4}}_{11,11}  (-1)^{a+b}   \nonumber\\
&&\hspace{-5.5cm} - \sum_{\nu_{3}\nu_{4}K}
 \Gamma_{K,ab}(E^{-}_{a,\nu_{1}\nu_{3}}) f_{K}(E^{-}_{a,\nu_{1}\nu_{3}}) 
X_{b,\nu_{4}\nu_{2}} X^{\dagger}_{a,\nu_{1}\nu_{3}}   \rho^{\nu_{3}\nu_{4}}_{00,00} 
   \nonumber\\
&&\hspace{-5.5cm} + \sum_{i\nu_{3}\nu_{4}K}
 \Gamma_{K,\overline{b}i}(E_{i,\nu_{4}\nu_{3}}^{+}) f_{K}(E_{i,\nu_{4}\nu_{3}}^{+}) 
X_{i,\nu_{3}\nu_{4}} X_{\overline{b},\nu_{4}\nu_{2}}^{\dagger}  
\rho^{\nu_{1}\nu_{3}}_{a,\overline{i}}  (-1)^{\overline{i}+b}   \nonumber\\
&&\hspace{-5.5cm} + \sum_{i\nu_{3}\nu_{4}K}
 \Gamma_{K,ib}(E^{-}_{i,\nu_{3}\nu_{4}}) (1-f_{K}(E^{-}_{i,\nu_{3}\nu_{4}})) 
 X_{b,\nu_{4}\nu_{2}}  X^{\dagger}_{i,\nu_{3}\nu_{4}} \rho^{\nu_{1}\nu_{3}}_{a,i} ,
    \nonumber
\end{eqnarray}
with
\begin{eqnarray}
E_{i\nu_{a}\nu_{b}}^{+} &=& \overline{\epsilon}_{i} + 
\overline{U}_{12} (1-\delta_{\overline{i}}) + \Omega(\nu_{a}-\nu_{b}). \nonumber
\end{eqnarray}
The respective $\langle 11 \vert..\vert 11 \rangle$ component reads
\begin{eqnarray}
\label{twostatesME0000III}
- 2 i \Omega (\nu_{1}-\nu_{2}) \rho_{11,11}^{\nu_{1}\nu_{2}} &=& \\
&&\hspace{-3.75cm} \phantom{+} \sum_{i\nu_{3}\nu_{4}K} 
\Gamma_{K,ii}(E_{i,\nu_{4}\nu_{3}}^{+}) (1-f_{K}(E_{i,\nu_{4}\nu_{3}}^{+})) 
 X_{i,\nu_{3}\nu_{4}} X^{\dagger}_{i,\nu_{1}\nu_{3}} \rho_{11,11}^{\nu_{4}\nu_{2}} 
\nonumber \\
&&\hspace{-3.75cm} - \sum_{ij\nu_{3}\nu_{4}K} 
\Gamma_{K,ij}(E_{j,\nu_{2}\nu_{4}}^{+})  f_{K}(E_{j,\nu_{2}\nu_{4}}^{+}) 
 X_{j,\nu_{4}\nu_{2}}  X^{\dagger}_{i,\nu_{1}\nu_{3}}
\rho_{\overline{i}\overline{j}}^{\nu_{3}\nu_{4}} (-1)^{i+j}  \nonumber\\
&&\hspace{-3.75cm} - \sum_{ij\nu_{3}\nu_{4}K} 
\Gamma_{K,ij}(E_{i,\nu_{1}\nu_{3}}^{+}) f_{K}(E_{i,\nu_{1}\nu_{3}}^{+}) 
 X_{j,\nu_{4}\nu_{2}} X^{\dagger}_{i,\nu_{1}\nu_{3}}  
\rho^{\nu_{3}\nu_{4}}_{\overline{i}\overline{j}} (-1)^{i+j}   \nonumber\\
&&\hspace{-3.75cm} + \sum_{i\nu_{3}\nu_{4}K} 
\Gamma_{K,ii}(E_{i,\nu_{3}\nu_{4}}^{+}) (1-f_{K}(E_{i,\nu_{3}\nu_{4}}^{+})) 
 X_{i,\nu_{4}\nu_{2}}  X^{\dagger}_{i,\nu_{3}\nu_{4}} \rho^{\nu_{1}\nu_{3}}_{11,11} .
   \nonumber
\end{eqnarray}
As for a single electronic state, this set of algebraic equations 
determines the density matrix describing transport through two electronic states. 
The current through two electronic states is explicitly given by
\begin{eqnarray}
\label{twostatecurrent}
\frac{1}{e} I_{K} &=& \\
&&\hspace{-1.75cm} -  \sum_{i,j,\nu_{1},\nu_{2},\nu_{3}}  
\left(1-f_{K}(E_{i,\nu_{2}\nu_{1}})\right) \Gamma_{K,ji}(E_{i,\nu_{2}\nu_{1}}) X_{i,\nu_{1}\nu_{2}}   
X_{j,\nu_{3}\nu_{1}}^{\dagger}  \rho_{ij}^{\nu_{2}\nu_{3}} \nonumber\\
&&\hspace{-1.75cm} -  \sum_{i,\nu_{1},\nu_{2},\nu_{3}}  
\left(1-f_{K}(E^{+}_{i,\nu_{2}\nu_{1}})\right) \Gamma_{K,ii}(E^{+}_{i,\nu_{2}\nu_{1}}) X_{i,\nu_{1}\nu_{2}}   
X_{i,\nu_{3}\nu_{1}}^{\dagger}  \rho^{\nu_{2}\nu_{3}}_{11,11} \nonumber\\
&&\hspace{-1.75cm} +  \sum_{i,j,\nu_{1},\nu_{2},\nu_{3}} \left( -1 \right)^{i+j} 
 f_{K}(E^{+}_{i,\nu_{1}\nu_{2}}) \Gamma_{K,ij}(E^{+}_{i,\nu_{1}\nu_{2}})  X_{j,\nu_{3}\nu_{1}} X^{\dagger}_{i,\nu_{1}\nu_{2}} 
 \rho_{\overline{i}\overline{j}}^{\nu_{2}\nu_{3}} \nonumber\\ 
&&\hspace{-1.75cm} +  \sum_{i,\nu_{1},\nu_{2},\nu_{3}} 
 f_{K}(E_{i,\nu_{1}\nu_{2}})  \Gamma_{K,ii}(E_{i,\nu_{1}\nu_{2}})  X_{i,\nu_{3}\nu_{1}}  X^{\dagger}_{i,\nu_{1}\nu_{2}}  
\rho^{\nu_{2}\nu_{3}}_{00,00} \nonumber\\ 
&&\hspace{-1.75cm} +  \sum_{i,\nu_{1},\nu_{2},\nu_{3}} 
 f_{K}(E_{i,\nu_{3}\nu_{2}}) \Gamma_{K,ii}(E_{i,\nu_{3}\nu_{2}}) 
X_{i,\nu_{2}\nu_{3}}  X_{i,\nu_{3}\nu_{1}}^{\dagger} \rho_{00,00}^{\nu_{1}\nu_{2}}  \nonumber\\ 
&&\hspace{-1.75cm} +  \sum_{i,j,\nu_{1},\nu_{2},\nu_{3}} \left( -1 \right)^{i+j} 
 f_{K}(E^{+}_{i,\nu_{3}\nu_{2}}) \Gamma_{K,ji}(E^{+}_{i,\nu_{3}\nu_{2}})  X_{i,\nu_{2}\nu_{3}}  X_{j,\nu_{3}\nu_{1}}^{\dagger} 
\rho_{\overline{j}\overline{i}}^{\nu_{1}\nu_{2}}
 \nonumber\\ 
&&\hspace{-1.75cm} -  \sum_{i,\nu_{1},\nu_{2},\nu_{3}} 
\left(1-f_{K}(E^{+}_{i,\nu_{2}\nu_{3}})\right) \Gamma_{K,ii}(E^{+}_{i,\nu_{2}\nu_{3}})   
 X_{i,\nu_{3}\nu_{1}} X_{i,\nu_{2}\nu_{3}}^{\dagger} \rho_{11,11}^{\nu_{1}\nu_{2}}  \nonumber\\
&&\hspace{-1.75cm} -  \sum_{i,j,\nu_{1},\nu_{2},\nu_{3}} 
 \left(1-f_{K}(E_{i,\nu_{2}\nu_{3}})\right) \Gamma_{K,ij}(E_{i,\nu_{2}\nu_{3}})   
 X_{j,\nu_{3}\nu_{1}} X_{i,\nu_{2}\nu_{3}}^{\dagger} \rho_{ji}^{\nu_{1}\nu_{2}}. \nonumber
\end{eqnarray}
Thereby, principal value terms, 
as for the computation of the reduced density matrix, are disregarded.

\section{Effects of electronic and vibrational coherences}
\label{effectofvibrationalcoherences}

None of the results that we have discussed in Sec.\ \ref{secResults} are significantly 
influenced by electronic or vibrational coherences of the reduced density matrix. 
This is due to the 
specific model parameters used. The respective eigenstates 
$\vert a\rangle\vert\nu\rangle$ ($a\in\lbrace 0,1 \rbrace$ or $a\in\lbrace 00,01,10,11 \rbrace$) 
do not exhibit any $\text{(quasi-)}$degeneracies. However, in realistic systems 
the molecular states $\vert a\rangle\vert\nu\rangle$ can show some (quasi-)degeneracy, 
and therefore, we investigate the effect of such coherences in this section. 

For a single electronic state and a single harmonic mode, the molecular states are quasi-degenerate, 
 if the broadening of the levels exceeds the level spacing set by the frequency $\Omega$. 
But in this regime, where $\Omega\lesssim\Gamma$, a perturbative expansion 
in $V_{k}$ may not be appropriate. Therefore, we start our discussion with a 
model system comprising two electronic states. We expect coherences to play a major 
role for degenerate or quasi-degenerate electronic states, \emph{i.e.}\ 
for $\vert\epsilon_{1}-\epsilon_{2}\vert<\Gamma$ \cite{Harbola2006}. 
As long as the couplings of the two quasi-degenerate electronic states to the leads are 
symmetric, we find the same results with and without 
coherences. This can be understood by inspection of 
Eqs.\ (\ref{twostatesME0000}) and (\ref{twostatesME0000III}). For symmetric junctions
with $\epsilon_{1}\approx\epsilon_{2}$ and with $\overline{U}\approx0$, 
where $\rho_{00,00}\approx\rho_{11,11}$ holds, Eqs.\ (\ref{twostatesME0000}) 
and (\ref{twostatesME0000III}) have almost the same structure. The only difference, in that case, 
is the sign by which coherences enter these equations. For that reason, the coherences 
$\rho_{1,2}=\rho_{2,1}^{*}$ cancel in these equations and, as a result, do not influence the 
respective transport characteristics. Only for non-symmetric couplings to the leads we 
find a significant effect of coherences on the transport characteristics of this system. 
The most pronounced effect appears, if one of the molecule-lead couplings differs by sign 
\emph{e.g.}\ for $\nu_{\text{L},1/2}=\nu_{\text{R},1}=-\nu_{\text{R},2}$ and if the two 
states are degenerate $\epsilon_{2}=\epsilon_{1}$. This specific model system can 
be identically transformed to two orthogonal states, which are not interacting with each 
other and which are coupled to one of the leads only (either left or right). Hence, the 
current through this system is zero for any bias voltage
\cite{Holleitner2001,Kubala2002}. However, if we disregard electronic coherences,
 we obtain a finite current that corresponds to the current of two states symmetrically 
coupled to the leads. Thus, for quasi-degenerate electronic states that are non-symmetrically 
coupled to the leads, electronic coherences must be accounted for to obtain physically correct results. 

\begin{figure}
\resizebox{\newwidth}{\newheight}{
\includegraphics{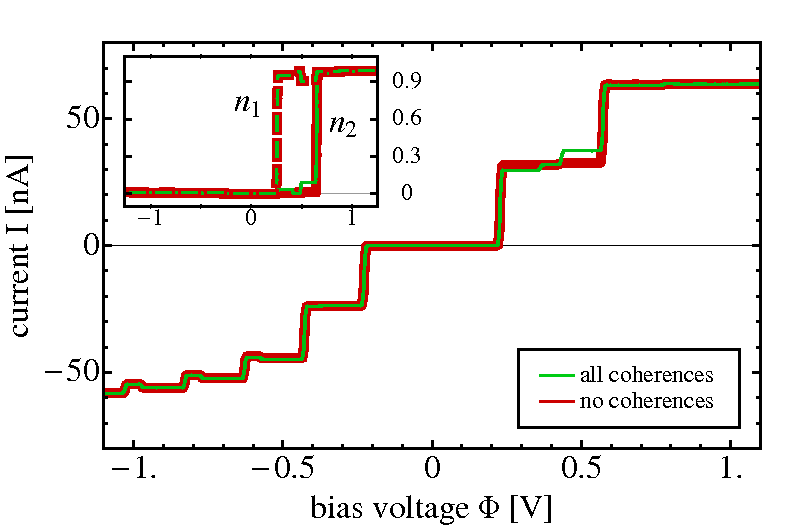}
}
\resizebox{\newwidth}{\newheight}{
\includegraphics{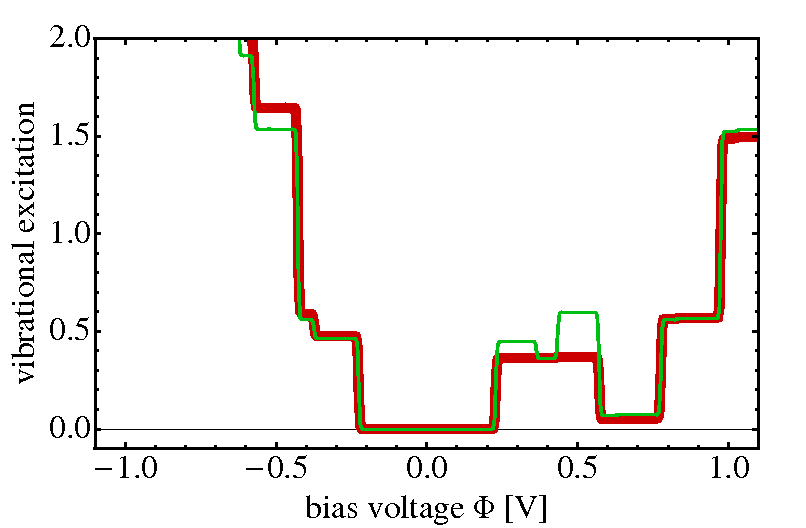}
}
\caption{\label{Coherences} Current-voltage characteristics and vibrational excitation 
for a model system similar to the one of Fig.\ \ref{ResAbAsym}. Here, the energy of 
the second electronic state is chosen such that the $\nu$th vibrational level of the 
electronically excited state of the anion is degenerate with respect to the $(\nu+1)$th 
vibrational state of the anionic ground-state: $\epsilon_{2}=\epsilon_{1}+\Omega$. 
The solid red line is obtained disregarding all coherences of the reduced density matrix, 
while the solid green line is obtained taking all coherences of $\rho$ into account. 
The inset shows the respective population of the electronic levels, where the dashed 
lines refer 
to the population of state 1 ($n_{1}$) and the solid lines to the one of state 2 ($n_{2}$). 
}
\end{figure}

The role of vibrational coherences can be studied employing a similar model system with 
two electronic states and energies that differ by the frequency of the vibrational mode, 
\emph{i.e.}\ $\epsilon_{2}=\epsilon_{1}+\Omega$. Systems with 
$\epsilon_{2}=\epsilon_{1}+n\Omega$ and $n\geq2$ display similar but less pronounced effects, 
as the impact of coherences decreases the further away they are located from the diagonal of 
the density matrix \cite{Harbola2006}. Again, for totally symmetric molecule-lead couplings 
$\nu_{K,i}$, we do not observe a significant influence of coherences. Only for 
asymmetric transport scenarios we find coherences to play a significant role for the 
transport characteristics. In Fig.\ \ref{Coherences} we present the current-voltage 
characteristics and the vibrational excitation for the asymmetric model system 
introduced in Sec.\ \ref{ResonantAbsorptionAsym} with the energy of the higher-lying 
state adjusted to $\epsilon_{2}=\epsilon_{1}+\Omega=0.25$\,eV. Thereby, the red line 
represents a calculation where all coherences are disregarded, while the green line 
represents the results for a calculation where all coherences are taken into account. 
Only in the vicinity $e\Phi=2\overline{\epsilon}_{1}$ to $e\Phi=2(\overline{\epsilon}_{2}+\overline{U})$ 
vibrational coherences influence the current-voltage characteristics. For positive bias voltages, 
the first step in the green line is diminished, since coherences result in a small 
population of the second higher-lying electronic state (cf.\ the inset of Fig.\ \ref{Coherences}a), 
which is thus blocking transport through the low-lying electronic state due to 
vibrationally induced repulsive electron-electron interactions $\overline{U}=-2\lambda_{1}\lambda_{2}/\Omega$. 
This blocking is lifted again for higher bias voltages, when electrons in the left 
lead have enough energy, \emph{i.e.}\ more 
than $\overline{\epsilon}_{1}+\overline{U}$. Vibrational excitation 
is enhanced as well, because there is an additional resonant emission process for 
tunneling from the higher-lying electronic state to the right lead.
Similarly, coherences result in a somewhat larger current for 
bias voltages $2\overline{\epsilon}_{2}<e\Phi<2(\overline{\epsilon}_{2}+\overline{U})$, 
where 
the higher-lying electronic state enters the bias window. 
Again, the population of the second electronic state is increased by vibrational coherences, 
although repulsive electron-electron interactions $\overline{U}$ block the population of, 
and similarly, transport through this state. Coherences soften this blocking, resulting in 
a larger current and vibrational excitation.

We expect systems with more than one vibrational degree of freedom to display gradually 
more quasi-degenerate levels, and hence, coherences to play a gradually more 
important role. Moreover for anharmonic potentials that are describing \emph{e.g.}\ molecular 
motors \cite{Pshenichnyuk2010}, coherences are crucial to characterize the actual 
motion of the molecule.


\begin{thebibliography}{125}
\bibitem[{not({\natexlab{d}})}]{noteonstrongRepulsion}
\bibinfo{note}{We like to note that even stronger Coulomb interactions
  $U>\vert\overline{\epsilon}_{2}-\overline{\epsilon}_{1}\vert$ result in a
  splitting of the di-anionic resonance, which thus appears at two different
  bias voltages: $e\Phi=2(\overline{\epsilon}_{1}+\overline{U})$ and
  $e\Phi=2(\overline{\epsilon}_{2}+\overline{U})$.}
\end{thebibliography}
\end{document}